\documentclass[12pt]{article}

\usepackage{epsfig}

\def \be  {\begin{equation}}
\def \ee  {\end{equation}}
\def \ba  {\begin{eqnarray}}
\def \ea  {\end{eqnarray}}
\def \baa {\begin{eqnarray*}}
\def \eaa {\end{eqnarray*}}

\def \MS {\overline{\rm MS}}

\def\hepph  #1 {{\tt hep-ph/#1}}

\begin{document}

\begin{flushright}
YITP-SB-02-49\\
\today
\end{flushright}

\vspace*{30mm}

\begin{center}
{\LARGE \bf Higher Orders in  $A\left(\alpha_s\right)/[1-x]_{+}$ of \\ \vspace*{2mm}
Non-Singlet Partonic Splitting Functions }

\par\vspace*{20mm}\par

{\large \bf
Carola F.\ Berger}

\bigskip

{\em C.N.\ Yang Institute for Theoretical Physics,
SUNY Stony Brook\\
Stony Brook, New York 11794 -- 3840, U.S.A.}

\end{center}
\vspace*{15mm}

\begin{abstract}
We develop a simplified method for obtaining higher orders in the perturbative expansion of the singular term  $A\left(\alpha_s\right)/[1-x]_{+}$ of non-singlet partonic splitting functions. Our method is based on the calculation of eikonal diagrams. The key point is the observation that the corresponding cross sections exponentiate in the case of two eikonal lines \cite{George, gath, freta}, and that the exponent is directly related to the functions $A\left(\alpha_s\right)$ due to the factorization properties of parton distribution functions. As examples, we rederive the one- and two-loop coefficients $A^{(1)}$ and $A^{(2)}$. We go on to derive the known general formula for the contribution  to $A^{(n)}$ proportional to $N_f^{n-1}$, where $N_f$ denotes the number of flavors. Finally, we determine the previously uncalculated term proportional to $N_f$ of the three-loop coefficient $A^{(3)}$ to illustrate the method. Our answer agrees with the existing numerical estimate \cite{vogt}.  The exact knowledge of the coefficients $A^{(n)}$ is important for the resummations of large logarithmic corrections due to soft radiation. Although only the singular part of the splitting functions is  calculable within our method, higher-order computations are much less complex than within conventional methods, and even the calculation of $A^{(4)}$ may be possible.
\end{abstract}

\newpage

\section{Introduction}

\subsection{Parton Distribution Functions}

Parton distribution functions (PDFs) are indispensable ingredients in any calculation within perturbative quantum chromodynamics involving hadrons in the initial state. $f_{a/A}(x)$ describes the distribution of parton $a$ in hadron $A$ with a momentum fraction $x$. It cannot be calculated within the framework of perturbation theory. However, its evolution is perturbatively calculable from the renormalization group equation \cite{DGLAP}
\be
\mu \frac{d}{d \mu} f_{a/A} (x,\mu) = \sum_b \int\limits_{x}^1 \frac{d \zeta}{\zeta} P_{a b}
\left(\zeta,\alpha_s(\mu)\right) f_{b/A} \left(\frac{x}{\zeta},\mu \right), \label{evol}
\ee
where $P_{a b}$ is the evolution kernel or splitting function, and $\mu$ denotes the factorization scale, usually taken equal to the renormalization scale. This follows from factorization \cite{pQCD}, which enables us to write a large class of  physical cross sections as convolutions of these
PDFs with perturbatively calculable short-distance functions.
Similar considerations apply to partonic cross sections. There the parton-in-parton distribution functions  $f_{f_i/f_j}$ describe the probability of finding parton $f_i$ in parton $f_j$. The evolutionary behavior of the partonic PDFs obeys the same equation, (\ref{evol}), as for the hadronic PDFs. Thus the splitting functions can be computed in perturbation theory.

The convolutions of these PDFs with short-distance functions simplify to simple products when we take moments:
\be
\tilde{f}_{f_1/f_2} (N,\mu) = \int_0^1 d x x^{N-1} f_{f_1/f_2} (x,\mu) = \int_0^\infty d x e^{-N(1-x)}
f_{f_1/f_2} (x,\mu) + {\mathcal{O}}\left(\frac{1}{N}\right). \label{moment}
\ee
The first definition in Eq. (\ref{moment}) is the Mellin transform which is equivalent to the second definition, the Laplace transform, at large $N$ because then $e^{-N(1-x)} \sim x^N$. Tildes here and below denote quantities in moment space.

Up to corrections of order $1/N$ we can neglect flavor mixing, that is, we deal with non-singlet distributions:
\be
f_{NS}^{\pm} = f^\pm_{q_a/q} - f^\pm_{q_b/q},\quad  f^\pm_{q_i/q} = f_{q_i/q} \pm f_{\bar{q}_i/q}.
\ee
The solution to the evolution equation (\ref{evol}) for non-singlet parton-in-parton distributions is in moment space given by
\begin{equation}
\tilde{f}_f (N, \mu, \varepsilon) = \exp \left[ \int_0^{\mu^2} \frac{d \mu'^2}{\mu'^2} \gamma_{ff} \left(N,\alpha_s(\mu') \right) \right] + {\mathcal{O}}\left(\frac{1}{N} \right), \label{exppdf}
\end{equation}
where $\gamma_{ff}(N) = \int_0^1 dx x^{N-1} P_{ff}(x)$ are the moments of the splitting functions, and  where we have imposed the boundary condition $\tilde{f}_f(N,\mu = 0,\varepsilon) = 1$. The argument $\varepsilon$ in $\tilde{f}_f$ indicates that we compute in the usual modified minimal subtraction scheme ($\MS$) in $n = 4 - 2 \varepsilon$ dimensions.

To leading power in $N$  the moments of the  partonic splitting functions,  $\gamma_{ff}(N)$, take the simple form \cite{korch,alball}
\begin{equation}
\gamma_{ff}(N,\alpha_s) =  A_f (\alpha_s) \ln N +  B_f (\alpha_s) + \mathcal{O}\left(\frac{1}{N} \right),\label{pffN}
\ee
or in $x$-space,
\begin{equation}
P_{ff} (x,\alpha_s ) = A_f (\alpha_s) \left[\frac{1}{1-x}\right]_{+} + B_f (\alpha_s) \delta(1-x) + {\mathcal{O}}\left(\left[1-x \right]^0 \right), \label{pff}
\end{equation}
where the plus distribution $\left[\frac{1}{1-x}\right]_{+}$ is defined by
\be
\int_z^1 dx f(x) \left[\frac{1}{1-x}\right]_{+} = \int_z^1 dx \frac{f(x)- f(1)}{1-x} + f(1) \ln (1-z).
\ee
The term with the plus distribution represents the cancellation of a single overall infrared divergence. The coefficient of $\ln N$ in Eq. (\ref{pffN}) can be expanded in the strong coupling,
\begin{equation}
A_f (\alpha_s) = \sum_n \left( \frac{ \alpha_s}{\pi} \right)^n A_f^{(n)}. \label{aaa}
\end{equation}

The exact knowledge of the terms $A^{(n)}$ is important for $x \rightarrow 1$ (large $N$), since there large logarithmic corrections arise due to soft-gluon radiation. These corrections need to be resummed in order to be able to make reliable predictions within perturbation theory. The knowledge of the coefficients $A^{(3)}$ and $B^{(2)}$ is required for threshold resummation \cite{threshold} at the next-to-next-to-leading logarithmic (NNLL) level \cite{vogt2}.

The anomalous dimensions $\gamma_{ff}(N)$ are currently known to two loops \cite{2loopknown}, and a general formula for the $\alpha_s^n N_f^{n-1}$-terms of $\gamma_{ff}(N)$ was computed by Gracey \cite{Gracey}. Work on the 3-loop splitting functions, based on the operator product expansion (OPE), is in progress \cite{3loop}. From the known exact values for some specific moments and the behavior at small $x$ \cite{moments} a numerical parametrization for the coefficient $A^{(3)}$ was obtained by Vogt \cite{vogt}, although, for the above reasons, the exact knowledge of this term is desirable. The above mentioned calculation by Moch, Vermaseren, and Vogt of the full three-loop splitting functions via the OPE in moment space will be completed in the near future \cite{MoVe}. Their results for the fermionic contributions are now available \cite{MVV}. However, the method presented here, although only applicable for the calculation of the coefficients $A$, not of the complete $x$-dependence, is complementary to the OPE method in two ways: we calculate only virtual diagrams, and furthermore, it is much less computationally intensive, thus a computation of the four- or even higher loop coefficients may be feasible.

In the following it is convenient to use light-cone coordinates, where our convention is as follows:
\ba
k^+ & = & \frac{1}{\sqrt{2}}\left(k^0 + k^3\right), \nonumber \\
k^- & = & \frac{1}{\sqrt{2}}\left(k^0 - k^3 \right), \\
k_\perp & =  & \left(k^1,k^2\right). \nonumber
\ea

Factorization allows us to define PDFs in terms of nonlocal operators.  At leading order in $N$ one can define \cite{pQCD, pdfs} the following function
\ba
f_{q_i/q} (x) & = & \frac{1}{4 \pi} \int d y^- e^{-i x p^+ y^-} \left< p \left| \bar{\psi}_i (0,y^-,0_\perp) \gamma^+ P e^{i g \int_0^{y^-} d z^- {\mathcal{A}}^{(q)\, +} (0,z^-,0_\perp)} \psi_i(0) \right| p \right> \nonumber \\
& = &  \frac{1}{4 \pi} \int d y^- e^{-i x p^+ y^-} \sum_n \left< p \left| \bar{\Psi}_i (0,y^-,0_\perp)\right| n \right>
\gamma^+ \left< n \left|  \Psi_i(0)  \right| p \right>, \label{pdfdef}
\ea
which describes the distribution of a quark $q_i$, created by the operator $\bar{\psi}_i$, in a quark $q$ with momentum $p$. The operators are separated by a light-like distance, and are joined with a path-ordered exponential, denoted by $P$, to achieve gauge-invariance. This exponential describes the emission of arbitrarily many  gluons of polarization in the plus-direction. ${\mathcal{A}}^{(q)}$ is the vector potential in the fundamental representation. In the second line we have inserted a complete set of states, have used the identity
\be
P e^{i g \int_0^{\eta} d \xi \beta \cdot {\mathcal{A}} (\xi n^\mu )}  = \left[P e^{i g \int_0^{\infty} d \xi \beta \cdot {\mathcal{A}} ( (\xi+\eta) \beta^\mu )}\right]^\dagger \left[ P e^{i g \int_0^{\infty} d \xi \beta \cdot {\mathcal{A}} (\xi \beta^\mu )} \right],  \label{phaseop}
\ee
and have defined
\be
\Psi_i(y) = P e^{i g \int_0^{\infty} d \xi \beta \cdot {\mathcal{A}} (y + \xi \beta^\mu )}  \psi_i(y),
\ee
with the light-like vector $\beta^\mu$ chosen in the minus-direction, $\beta^+ = \beta_\perp = 0$.

The Feynman rules for the expansion of an ordered exponential are shown in Fig. \ref{Frules}, where the double lines represent ``eikonal'' propagators. Thus, path-ordered exponentials are closely connected to the eikonal, or soft approximation, which results in the same Feynman rules, as we will see below. Gluon parton distribution functions can be constructed analogously \cite{pQCD, pdfs}, with the vector potentials in the adjoint representation, and appropriate operators for the creation of gluons.

\begin{center}
\begin{figure}[htb]
\begin{center}
\epsfig{file=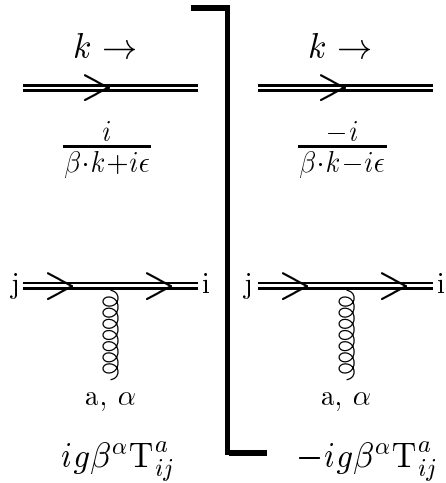,height=7.5cm,clip=0}
\caption{
Feynman rules for eikonal lines in the fundamental representation with velocities $\beta^\mu$, represented by the double lines. The vertical line represents the cut separating the amplitude and its complex conjugate. For an eikonal line in the adjoint representation one has to replace $T^a_{ij}$ with $i f_{ija}$. \label{Frules} }
\end{center}
\end{figure}
\end{center}

\subsection{Outline of the Paper}

Below we will outline a method which greatly simplifies the calculations of the collinear coefficients $A^{(n)}$ in the non-singlet case. Our method is based on the factorized form of the partonic non-singlet PDFs at leading order, where for $x \rightarrow 1$ the collinear coefficients are factored into an eikonal cross section. It was observed by Sterman that these cross sections exponentiate \cite{George}, not only in abelian theories, but also in nonabelian theories. As we will see, this exponentiation further simplifies the technical calculations, and enables us to go a step further towards the full calculation of the coefficient $A^{(3)}$. To illustrate the method, we compute the as yet undetermined contribution proportional $N_f$ to the three-loop coefficient.

A similar observation was made by Korchemsky \cite{korch}, who related the anomalous dimension of PDFs, Eq. (\ref{pff}), with the cusp anomalous dimension of a Wilson loop. His work was performed in a noncovariant axial gauge, whereas here  we will use Feynman gauge throughout. Korchemsky's observation was used in \cite{korchmarch} for the calculation of the two-loop coefficient $A^{(2)}$, which was done in Feynman gauge.  The work of Ref. \cite{korchmarch} was also based on the renormalization properties and exponentiation of Wilson loops (see \cite{cusp} and references therein). This approach is related to ours. However, the additional observations we make below result in several advantages.  The number of diagrams contributing at each order is decreased by working with light-like eikonals. Furthermore, as we will show below, we can restrict ourselves only to virtual graphs. With the help of Ward identities we are also able to justify the eikonal approximation and to show explicitly the absence of infrared (IR) subdivergences and the cancellation of ultraviolet (UV) subdivergences at the eikonal vertex, leaving only the usual QCD UV divergences.

The discussion consists of two parts. In the first part, Sections \ref{sectwebA} and \ref{sectexp}, we lay the theoretical foundation of our method, whose technical details are described and explained by means of various examples in the second part, Sections \ref{sectmeth} and \ref{sect3loop}.

We will first derive in Sect. \ref{sectwebA} a factorized form for PDFs, and introduce an eikonal factor that absorbs all collinear singularities. This implies  that  the $A(\alpha_s)$ can be derived from the eikonal factor. In the next section we will review the proof of the exponentiation \cite{gath,freta} of this factor, show how to construct the exponent order by order in perturbation theory, and prove the absence of infrared/collinear subdivergences at each order. These properties then allow us to develop an algorithm, using light-cone ordered perturbation theory (LCOPT), for the systematic calculation of the collinear coefficients $A$, which is equivalent to the calculation of the virtual contribution to the anomalous dimension of an eikonal vertex. We summarize the method in Section \ref{sectmeth},  before going on to rederive as examples the one- and two-loop coefficients $A^{(1)}$ and $A^{(2)}$. In Section \ref{sect3loop} we derive a formula for the coefficients of $A^{(n)}$ proportional to $N_f^{n-1}$, which agrees with the corresponding contribution computed by Gracey \cite{Gracey} using an effective theory. We end by illustrating the steps necessary for the complete calculation of the 3-loop coefficient $A^{(3)}$. The IR structure of $A^{(3)}$ is explored for the graphs contributing at $\alpha_s^3 N_f$, which we calculate exactly and compare to Vogt's \cite{vogt} parametrization. In the Appendix we list the Feynman rules for LCOPT.

We will work in Feynman gauge, which further simplifies the expressions and reduces the number of diagrams, since then self-energies of light-like eikonal lines vanish.
All our explicit calculations are performed in the $\MS$ scheme, but our method is applicable in any minimal subtraction scheme.

\section{Non-Singlet Parton Distribution Functions and Eikonal Cross Sections} \label{sectwebA}

Here we will show that the factorized form of a perturbative non-singlet parton-in-parton distribution function contains an eikonal cross section which absorbs all collinear and infrared singular behavior as $x \rightarrow 1$.

\subsection{Leading Regions} \label{powercount}

We start with the definition Eq. (\ref{pdfdef}) of a perturbative parton-in-parton distribution function, which is shown in Fig. \ref{partondef} a) in cut-diagram notation. The part to the left of the cut (the vertical line) represents the amplitude, whereas the complex conjugate amplitude is drawn to the right of the cut.  We pick the incoming momentum to flow in the plus direction, $p^\mu = (p^+,0,0_\perp)$. Because the minus and transverse momenta in (\ref{pdfdef}) are integrated over, they can flow freely through the eikonal line, whereas no plus momentum flows across the cut in the eikonal line.

\begin{figure}
\begin{center}
\epsfig{file=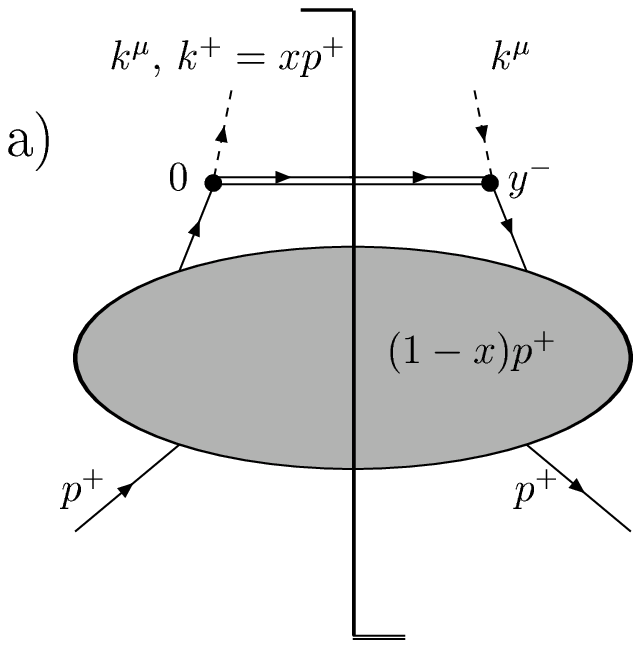,height=5.3cm,clip=0}
\vspace*{2mm} \\
\epsfig{file=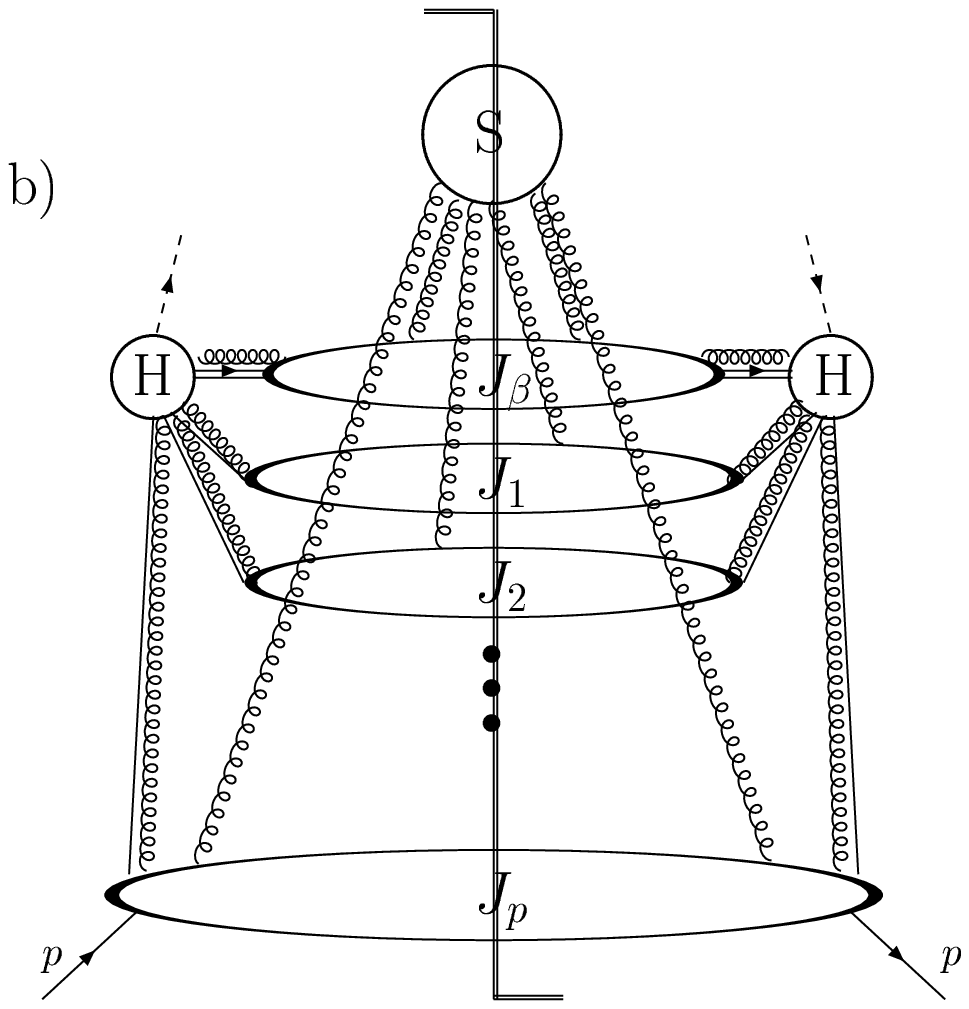,height=7.1cm,clip=0}
\\
\vspace*{1mm}
\epsfig{file=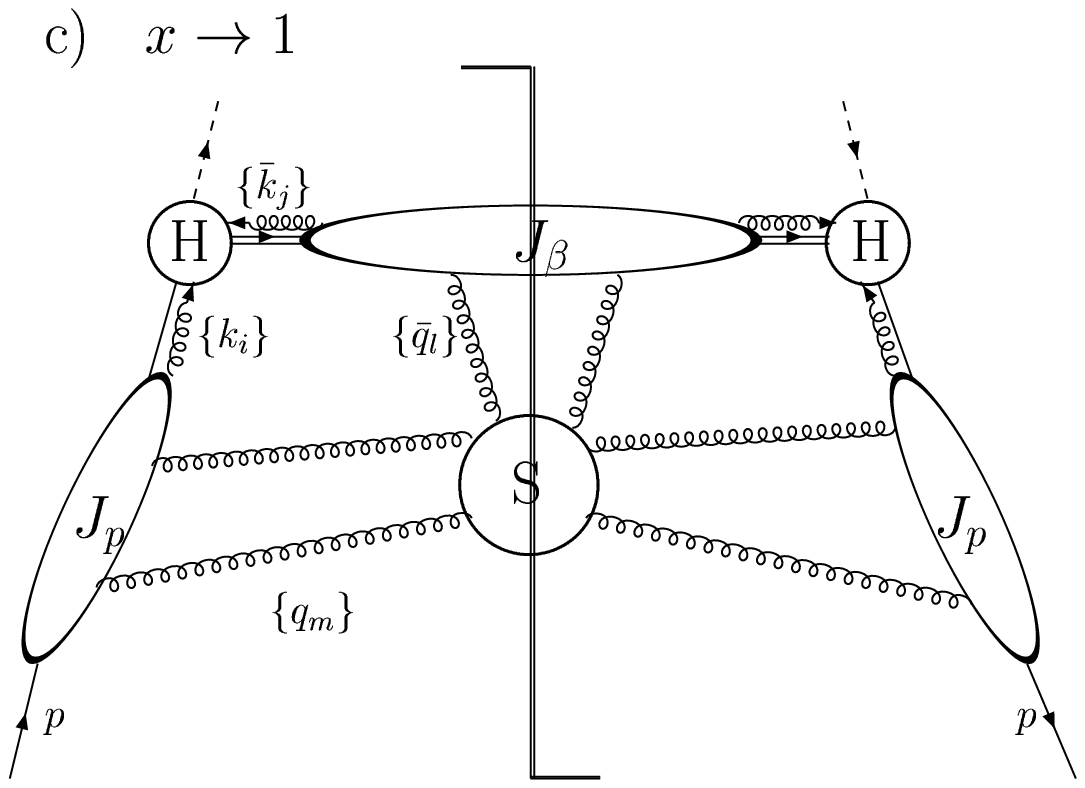,height=6.2cm,clip=0}
\caption{a) Graphical representation of a parton-in-parton distribution function, Eq. (\ref{pdfdef}), and b) its factorized form for arbitrary momentum fraction $x$, drawn as a reduced diagram. c) The reduced diagram for the PDF in the limit $x \rightarrow 1$. Here we have suppressed the labels $L$ and $R$  for purely virtual contributions to the left and to the right of the cut, respectively, compared to the notation in the text. The cut represents the final state.} \label{partondef}
\end{center}
\end{figure}

We follow the method developed in \cite{bookTASI, count} to find the regions in momentum space which can potentially contribute to the leading infrared singular behavior. These regions are shown in Fig. \ref{partondef} b) in form of a reduced diagram where all off-shell momenta are contracted. Momentum regions which can contain singularities are found with the help of the Landau equations \cite{landau}, which identify the singularities in Feynman integrals. However, not all regions which satisfy the Landau equations give leading contributions. The leading contributions are found by  infrared power counting \cite{count}.

 We find the leading  regions for the PDF which
 are associated with the following momentum configurations, scaled relative to  the large momentum scale in the problem, $x p^+ \approx p^+$:
\begin{itemize}
\item Soft momenta, which scale as $k^\mu \sim \lambda\, p^+, \, \lambda \ll 1$, in all components. The corresponding lines are denoted by $S$ in the figure.
\item Soft momenta with components which scale in the strongly ordered form $k^+ \sim \sigma \, p^+, k^- \sim \lambda \, p^+, k_\perp^2 \sim \lambda \, (p^+)^2$, or  $k^- \sim \sigma\, p^+, k^+ \sim  \lambda\, p^+, k_\perp^2 \sim \lambda\, (p^+)^2$, respectively, where $\sigma \ll \lambda \ll 1$. These are the so-called Glauber- or Coulomb momenta \cite{glauber, coste}.
\item Momenta collinear to the momenta of initial or final state particles (including the eikonal). The corresponding subdiagrams are denoted as $J_m$, where $m$ labels the external momentum. Momenta collinear to particles moving in the plus direction scale as $k^+ \sim  p^+, k^- \sim \lambda\, p^+, k_\perp^2 \sim \lambda\, (p^+)^2$, whereas momenta collinear to the minus direction behave as $k^- \sim  p^+, k^+ \sim\lambda\, p^+, k_\perp^2 \sim \lambda\, (p^+)^2$.
\item These regions are connected by a hard scattering, $H$, where all momenta are far off-shell, and thus scale as $\sim \, p^+$ in all components.
\end{itemize}

In our case we can have a jet $J_p$ collinear to the incoming momentum $p^+$, as well as an arbitrary number of jets $J_i$ emerging from the hard scattering. Furthermore, we can have momenta collinear to the eikonal moving in the minus direction, $\beta^-$, represented by $J_\beta$ in the figure. The jets can be connected by arbitrarily many soft gluons, $S$. Here and below, unless explicitly stated otherwise, we use the term ``soft'' for both soft and Glauber momenta.

A further simplification of the leading behavior occurs when we perform the limit to $x \rightarrow 1$, in which we are interested here. Jet lines having a finite amount of plus momentum in the final state become soft, and only virtual contributions can have large plus momenta. This does not affect the jet collinear to the eikonal, since it is moving in the minus direction. Thus we arrive at the leading regions depicted in Fig. \ref{partondef} c), with hard scatterings $H_L$ and $H_R$, which are the only vertices where finite amounts of momentum can be transferred, virtual jets $J_{p,\,L}$ and $J_{p,\,R}$ collinear to the incoming momentum $p^+$, a jet $J_\beta$ collinear to the eikonal $\beta^-$, connected via soft momenta, $S$. Here and below the subscripts $L$ and $R$, respectively, indicate that the momenta and functions are purely virtual, located to the left or to the right of the cut.

Infrared power counting \cite{bookTASI, count} shows now that the parton distribution function for $x \rightarrow 1$, as shown in Fig. \ref{partondef} c), can at most diverge as one inverse power, which corresponds to the $\frac{1}{1-x}$ divergence we are expecting from Eq. (\ref{pff}). Since the power counting is straightforward we just state the result here. To get the maximum degree of IR divergence the following conditions have to be fulfilled (see the figure):
\begin{itemize}
\item No soft vectors directly attach $H_L$ or $H_R$ with $S$.
\item The jets and the soft part can only be connected through soft gluons, denoted by the sets $\left\{q_{L,\,m_L}\right\}$, $\left\{q_{R,\,m_R}\right\}$, and $\left\{\bar{q}_l\right\}$.
\item Arbitrarily many scalar-polarized gluons, $\left\{k_{L,\,i_L}\right\}$, $\left\{k_{R,\,i_R}\right\}$, $\{\bar{k}_{L,\,j_L}\}$ and $\{\bar{k}_{R,\,j_R}\}$ , attach the jets with $H_L$ and $H_R$, respectively. The barred momenta are associated with $J_\beta$, the unbarred with the $J_p$'s.
\item Exactly one scalar, fermion, or physically polarized gluon with momentum $k_L^\mu - \sum\limits_{i_L} k_{L,\,i_L}^\mu$, $k_R^\mu - \sum\limits_{i_R} k_{R,\,i_R}^\mu$,  $\bar{k}_L^\mu - \sum\limits_{j_L} \bar{k}_{L,\,j_L}^\mu$ or  $\bar{k}_R^\mu - \sum\limits_{j_R} \bar{k}_{R,\,j_R}^\mu$, respectively, connects each of the jets with the hard parts.
The momenta $k_L^{\nu_L}$ ($k_R^{\nu_R}$) and $\bar{k}_L^{\rho_L}$ ($\bar{k}_R^{\rho_R}$) denote the \emph{total} momenta flowing into $H_L$ ($H_R$) from $J_{p,\,L}$ ($J_{p,\,R}$) and $J_\beta$, respectively, that is, they are the sum of the scalar, fermion or physically polarized gluon momenta, and the scalar-polarized gluon momenta.

In an individual diagram we can have only scalar-polarized gluons connecting the jets with the hard parts, and no scalar, fermion or physically polarized gluon. However, the sum of these configurations vanishes after application of the Ward identity shown in Fig. \ref{wardeik} a).
\item The number of soft and scalar-polarized vector lines emerging from a particular jet is less or equal to the number of 3-point vertices in that jet.
\end{itemize}

In summary, we have found that the regions in momentum space which give leading contributions may be represented as:
\ba
f_{f}^{x \rightarrow 1}\;(x) &\! = & \! \sum\limits_{C_\beta,C_S} \int\frac{d^n k_L}{(2 \pi)^n} \int \frac{d^n k_R}{(2 \pi)^n}\int\frac{d^n \bar{k}_L}{(2 \pi)^n}  \int\frac{d^n \bar{k}_R}{(2 \pi)^n} \prod\limits_{i_L,i_R,j_L,j_R} \int \frac{d^n k_{L,\,i_L}}{(2 \pi)^n}  \frac{d^n k_{R,\,i_R}}{(2 \pi)^n}  \frac{d^n \bar{k}_{L,\,j_L}}{(2 \pi)^n}  \frac{d^n \bar{k}_{R,\,j_R}}{(2 \pi)^n}\nonumber \\
& &  \prod\limits_{m_L,m_R,l} \int \frac{d^n q_{L,\,m_L}}{(2 \pi)^n} \frac{d^n q_{R,\,m_R}}{(2 \pi)^n} \frac{d^n \bar{q}_l}{(2 \pi)^n} \;  S^{(C_S)} \left( \{q_{L,\,{m_L}}^{\gamma_{k_L}} \}; \{q_{R,\,{m_R}}^{\gamma_{k_R}} \};\{\bar{q}_l^{\delta_l}\} \right) \nonumber \\
& & \,\, \times \;  H_L \left( k_L^{\nu_L} , \{k_{L,\,i_L}^{\alpha_{i_L}} \}; \bar{k}_L^{\rho_L}, \{\bar{k}_{L,\,j_L}^{\eta_{j_L}} \} \right)\;
H_R \left( k_R^{\nu_R} , \{k_{R,\,i_R}^{\alpha_{i_R}} \}; \bar{k}_R^{\rho_R}, \{\bar{k}_{R,\,j_R}^{\eta_{j_R}} \} \right) \nonumber \\
& & \,\, \times \;  J_{p,\,L} \left(k_L^{\nu_L}, \{k_{L,\,i_L}^{\alpha_{i_L}} \};  \{q_{L,\,{m_L}}^{\gamma_{m_L}} \} \right) \;  J_{p,\,R} \left(k_R^{\nu_R}, \{k_{R,\,i_R}^{\alpha_{i_R}} \};  \{q_{R,\,{m_R}}^{\gamma_{m_R}} \} \right) \nonumber \\
& & \,\, \times \; J_\beta^{(C_\beta)} \left(\bar{k}_L^{\rho_L}, \{\bar{k}_{L,\,j_L}^{\eta_{j_L}} \}; \bar{k}_R^{\rho_R}, \{\bar{k}_{R,\,j_R}^{\eta_{j_R}} \};\{\bar{q}_l^{\delta_l}\} \right)   \label{nonfactor} \\
& &\,\, \times \; \delta^n\left(\sum_{m_L} q^\mu_{L,\,m_L} + \sum_{m_R} q^\mu_{R,\,m_R} +  \sum_l \bar{q}^\mu_{l} \right) \;  \delta^n\left(k^\mu_L + \bar{k}^\mu_L - x p^\mu \right) \;\delta^n\left(k^\tau_R + \bar{k}^\tau_R - x p^\tau \right) \; \nonumber \\
& & \,\, \times \;  \delta^n\left( k_L^\mu -\sum_{m_L} q^\mu_{L,\,m_L} - p^\mu \right) \; \delta^n\left( k_R^\tau - \sum_{m_R} q^\tau_{R,\,m_R}  - p^\tau \right)\;. \nonumber
\ea
The sum in Eq. (\ref{nonfactor}) runs over all cuts of jet $J_\beta$, $C_\beta$,  and of the soft function $S$, $C_S$, which are consistent with the constraints from the delta-functions due to momentum conservation.
The functions in (\ref{nonfactor}) are still connected with each other by scalar polarized or soft gluons. This obscures the independent evolution of the functions. In the next subsection we will show how to simplify this result.

\subsection{Factorized Form of the Perturbative Distribution Function}

Following \cite{pQCD}, we will show here that the scalar polarized gluons decouple via the help of a Ward identity, and that we can disentangle the jets and the soft part, which are connected by soft gluon exchanges, via the so-called soft approximation.

\subsubsection{Decoupling of the Hard Part}

Starting from Eq. (\ref{nonfactor}), we use the fact that the leading contributions come from regions where the gluons carrying momenta $\left\{k_{L,\,{i_L}}\right\}$, $\left\{k_{R,\,{i_R}}\right\}$, $\{\bar{k}_{L,\,{j_L}}\}$ and $\{\bar{k}_{R,\,{j_R}}\}$ are scalar polarized. Thus $H_L$, $H_R$, $J_{p,\,L}$, $J_{p,\,R}$, and $J_\beta$ have the following structure:
\ba
H_L \! & = &\! \left( \prod\limits_{i_L} \xi^{\nu_{i_L}} \right) \left( \prod\limits_{j_L} \beta^{\rho_{j_L}} \right)
  H_{L,\, \{\nu_{i_L} \},\{\rho_{j_L}\} } \left(k_L^+\; \xi^{\nu_L}, \left\{k_{L,\,i_L}^+ \xi^{\alpha_{i_L}}\right\}; \bar{k}_L^- \; \beta^{\rho_L}, \{\bar{k}_{L,\,{i_L}}^- \beta^{\eta_{j_L}} \} \right),
 \\
J_{p,\,L} & = & \left( \prod\limits_{i_L} \beta^{\nu_{i_L}} \right)  J_{p,\,L \, \{\nu_{i_L} \} } \left(  k_L^{\nu_L}, \{k_{L,\,i_L}^{\alpha_{i_L}} \}; \{q_{L,\,{k_L}}^{\gamma_{m_L}} \} \right),
\\
J_\beta & = & \left( \prod\limits_{j_L} \xi^{\rho_{j_L}} \right) \left( \prod\limits_{j_R} \xi^{\rho_{j_R}} \right)
J_{\beta\, \{\rho_{j_L}\},\,\{\rho_{j_R}\}} \left(  \bar{k}_L^{\rho_L}, \{\bar{k}_{L,\,j_L}^{\eta_{j_L}} \};  \bar{k}_R^{\rho_R}, \{\bar{k}_{R,\,j_R}^{\eta_{j_R}} \} ;\{\bar{q}_l^{\delta_l}\}\right),
\ea
where, as above, for the functions to the right of the cut, we replace the subscripts $L$ with $R$. In these relations the vectors
\ba
\xi^\mu & = & \delta_+^\mu \, , \nonumber \\
\beta^\mu & = & \delta_{-}^\mu \,  \label{lightdef}
\ea
are the light-like vectors parallel to $p^\mu$ and parallel to the direction of the eikonal, respectively.

We now use the identity depicted in Fig. \ref{wardeik} c), where the grey blob denotes the hard part. Fig. \ref{wardeik} c) follows from the Ward identity shown in Fig. \ref{wardeik} a), and the identity for scalar polarized gluons attaching to an eikonal line in Fig. \ref{wardeik} b). The Ward identity says that the sum of all possible attachments of a scalar-polarized gluon to a matrix element vanishes. From this follows Fig. \ref{wardeik} a), since, by definition, we do not include the graph into the hard function where the gluon attaches to the physically polarized parton (quark or gluon), shown on the right-hand side of Fig. \ref{wardeik} a). The eikonal identity in Fig. \ref{wardeik} b) follows trivially from the eikonal Feynman rules in Fig. \ref{Frules}. Thus, since the right-hand sides of figures a) and b) are the same (the ``empty'' eikonal line carries no momentum), the left-hand sides are the same. Repeated application of this identity results in Fig. \ref{wardeik} c) \cite{CSS}. Note that the color factors are included in the Ward identity, resulting in the appropriate color factor for the attachments of the gluons as shown in Fig. c).

\begin{figure}[htb]
\begin{center}

\epsfig{file=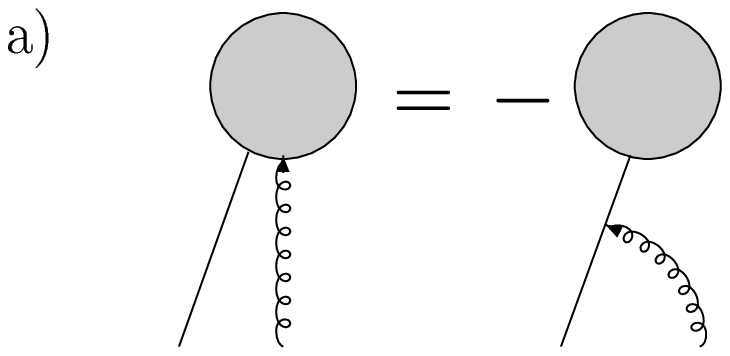,height=3.0cm,clip=0}
\hspace*{1cm}
\epsfig{file=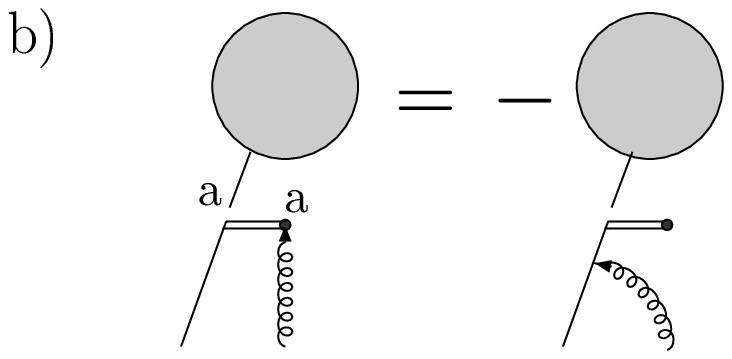,height=3.0cm,clip=0} \\
\vspace*{6mm}
\epsfig{file=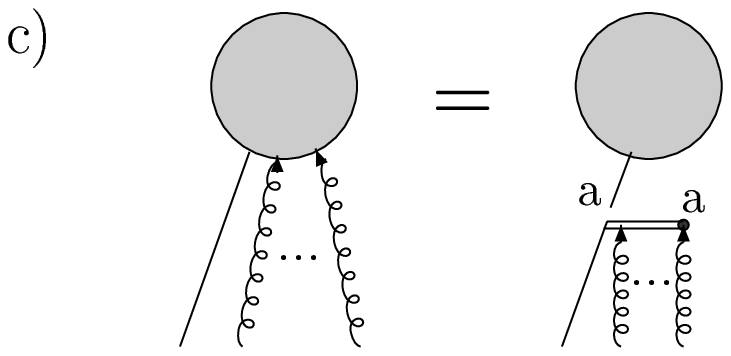,height=3.0cm,clip=0}
\caption{a) Ward identity for a scalar polarized gluon. b) Identity for a single longitudinally polarized gluon attaching to an eikonal line. c) Resulting identity after iterative application of Figs. a) and b). Repeated gauge-group indices are summed over. }  \label{wardeik}
\end{center}
\end{figure}

Using the identity in Fig. \ref{wardeik} c) for all scalar polarized gluons in the sets $\left\{k_{L,\,{i_L}}\right\}$, $\left\{k_{R,\,{i_R}}\right\}$, $\{\bar{k}_{L,\,{j_L}}\}$ and $\{\bar{k}_{R,\,{j_R}}\}$, we achieve our goal of decoupling the jet functions from the hard function. This decoupling occurs in Feynman gauge only after summation over the full gauge-invariant set of graphs which contribute to the reduced diagram Fig. \ref{partondef} c). The result is shown in Fig. \ref{partonfact1}. The products over the vectors $\beta$ and $\xi$ are replaced by eikonal factors $\mathcal{E}$, which we can group with the jets. Furthermore, the hard scatterings, by definition far off-shell, become  independent of $x$ up to corrections which vanish for $x \rightarrow 1$. Eq. (\ref{nonfactor}) then becomes
\ba
f_{f}^{x\rightarrow 1}\;(x) &\! = & \! H_L(p,\mu;\beta,\xi)\; H_R(p,\mu;\beta,\xi) \; \sum\limits_{C_\beta,C_S}  \prod\limits_{m_L,m_R,l}\int \frac{d^n q_{L,\,m_L}}{(2 \pi)^n} \frac{d^n q_{R,\,m_R}}{(2 \pi)^n} \frac{d^n \bar{q}_l}{(2 \pi)^n} \;
  \nonumber \\
& \times & \;  \tilde{J}_{p,\,L}(p,\mu;\beta; \{q_{L,\,{m_L}}^{\gamma_{m_L}} \} ) \; \tilde{J}_{p,\,R}(p,\mu;\beta; \{q_{R,\,{m_R}}^{\gamma_{m_R}} \})
 \;  \nonumber \\
& \times & \; \int dy \int dz\; S^{(C_S)} \left( y p,\mu; \{q_{L,\,{m_L}}^{\gamma_{m_L}} \}; \{q_{R,\,{m_R}}^{\gamma_{m_R}} \};\{\bar{q}_l^{\delta_l}\} \right) \; \tilde{J}^{(C_\beta)}_{\beta}(z p, \mu;\xi;\{\bar{q}_l^{\delta_l}\}) \;
  \label{factform1}\\
&  \times & \; \delta^n\left(\sum_{m_L} q^\mu_{L,\,m_L} + \sum_{m_R} q^\mu_{R,\,m_R} +  \sum_l \bar{q}^\mu_{l} \right) \; \delta(1-x-y-z) \;, \nonumber
\ea
where $\mu$ denotes the renormalization scale, which we set equal to the factorization scale, for simplicity. Corrections are subleading by a power of $1-x$. We define the functions $\tilde{J}_{p,\,L}$, $\tilde{J}_{p,\,R}$, and $\tilde{J}_\beta$ as follows:
\ba
\tilde{J}_{p,\,L} & = & \int\frac{d^n k_L}{(2 \pi)^n} \prod\limits_{i_L}  \int \frac{d^n k_{L,\,i_L}}{(2 \pi)^n}\; \mathcal{E} \left(\beta, \{k_{L,\,i_L}^+\} \right)^{\{\nu_{i_L}\}} \; J_{p,\,L \, \{\nu_{i_L} \} } \left(  k_L^{\nu_L}, \{k_{L,\,i_L}^{\alpha_{i_L}} \}; \{q_{L,\,{k_L}}^{\gamma_{m_L}} \} \right) \; \nonumber \\
& & \; \times \; \delta^n \left(k_L^\mu - \sum_{m_K} q^\mu_{L,\,m_L} - p^\mu\right), \label{Jp}  \\
\tilde{J}^{(C_\beta)}_\beta & = & \int\frac{d^n \bar{k}_L}{(2 \pi)^n}  \int\frac{d^n \bar{k}_R}{(2 \pi)^n}  \prod\limits_{j_L,j_R} \int \frac{d^n \bar{k}_{L,\,j_L}}{(2 \pi)^n} \frac{d^n \bar{k}_{R,\,j_R}}{(2 \pi)^n} \; \mathcal{E} \left(\xi, \{\bar{k}_{L,\,j_L}^-\} \right)^{\{\rho_{j_L}\}} \mathcal{E}^{*} \left(\xi, \{\bar{k}_{R,\,j_R}^-\} \right)^{\{\rho_{j_R}\}} \nonumber
 \\
& & \; \times \; J^{(C_\beta)}_{\beta\, \{\rho_{j_L}\},\,\{\rho_{j_R}\}} \left( zp; \bar{k}_L^{\rho_L}, \{\bar{k}_{L,\,j_L}^{\eta_{j_L}} \}; \bar{k}_R^{\rho_R}, \{\bar{k}_{R,\,j_R}^{\eta_{j_R}} \} ; \{\bar{q}_l^{\delta_l}\}\right) \;,
\ea
and $\tilde{J}_{p,\,R}$ is defined analogously to $\tilde{J}_{p,\,L}$, with the subscripts $L$ replaced by $R$, and with a complex conjugate eikonal line, since it is to the right of the cut. In (\ref{factform1}) the total plus momentum flowing across the cut is restricted to be $(1-x) p^+$, and flows through the soft function and/or the eikonal jet. The plus momenta flowing across the cuts $C_S$ and $C_\beta$, denoted by $y p^+$ and $z p^+$,  respectively, are therefore restricted to be $(1-x) p^+$ via the delta-function in Eq. (\ref{factform1}). Above we have factorized the hard part from the remaining functions, which are still linked via soft momenta.

\begin{figure}
\begin{center}
\epsfig{file=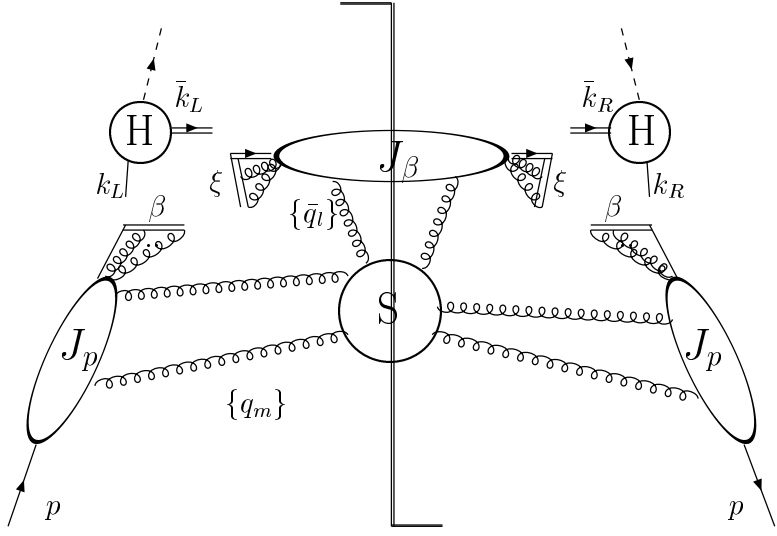,height=7.0cm,clip=0}
\caption{Parton distribution function for $x \rightarrow 1$ with the jet functions factorized from the hard part, graphical representation of Eq. (\ref{factform1}).} \label{partonfact1}
\end{center}
\end{figure}

\subsubsection{Fully Factorized Form} \label{softapprox}

Here we will use the soft approximation \cite{coste,CSS,back,pQCD} to factorize the jets $J_p$ from the soft function, which, by power counting, are connected only through soft gluons. We could factorize the eikonal jet $J_\beta$ from $S$ in an analogous way, but we choose not to do so here because eventually we will combine all soft and eikonal functions to form an eikonal cross section.

The soft approximation consists of two approximations: the neglect of non-scalar gluon polarizations, and the neglect of soft momenta $q^\mu$ in the numerator compared to jet-momenta $k^\mu$ and $k^2$ compared to $q \cdot k$ in the denominator. After making these approximations we proceed in a fashion similar to the previous subsection.

Let us start by decomposing each soft gluon propagator $D_{\mu \nu}$ in the sets $\left\{q_{L,\,m_L}\right\}$ and $\left\{q_{R,\,m_R}\right\}$ into a scalar-polarized contribution and a remainder \cite{GraYen}:
\ba
D_{\mu \nu} \left(q_{L,\,m_L}\right) & = & G_{\mu \nu} \left(q_{L,\,m_L},\xi \right) + K_{\mu \nu} \left(q_{L,\,m_L},\xi \right),  \label{kgdec}
\ea
and analogously for the momenta $\left\{q_{R,\,m_R}\right\}$ to the right of the cut, where we define
\ba
K_{\mu \nu} \left(q_{L,\,m_L},\xi \right) & \equiv & D_{\mu \rho} \left(q_{L,\,m_L}\right) \frac{\xi^\rho \, q_{L,\,m_L,\,\nu}}{\xi \cdot q_{L,\,m_L} + i \epsilon}, \nonumber \\
G_{\mu \nu} \left(q_{L,\,m_L},\xi \right) & \equiv & D_{\mu \nu} \left(q_{L,\,m_L}\right) - D_{\mu \rho} \left(q_{L,\,m_L}\right) \frac{\xi^\rho \, q_{L,\,m_L,\,\nu}}{\xi \cdot q_{L,\,m_L} + i \epsilon}, \label{kgdef}
\ea
 The sign of the $i \epsilon$-prescription is chosen in such a way as to not introduce new pinch singularities near the ones produced by the soft gluons under consideration. For momenta to the right of the cut in initial-state jets we also have $+i\epsilon$, since the sign of the momentum flow to the right of the cut is reversed relative to the jet momentum $k$. That is, we have for momenta to the left of the cut $(q+k)^2 + i \epsilon \approx 2 q \cdot k + i \epsilon$, whereas to the right of the cut we obtain $(q-k)^2 - i \epsilon \approx -2 q \cdot k - i \epsilon$.

From Eq. (\ref{kgdef}) we see, that for the scalar-polarized $K$-gluons the identity shown in Fig. \ref{wardeik} c) is immediately applicable, leading to the desired factorized form. So it remains to be shown that the $G$-gluons do not give leading contributions. As mentioned in the previous subsection  power-counting shows that only 3-point vertices are relevant for the coupling of soft gluons to jets. Let us consider a 3-point vertex, where a soft $G$-gluon with propagator $G_{\mu \nu}(q)$ couples to a fermion jet-line with momentum $k^\mu$ in the jet $J_p$ moving in the plus-direction:
\be
\frac{\not \! k + \not \! q}{(k+q)^2+i \epsilon} \gamma^\nu \frac{\not \! k}{k^2+i\epsilon} G_{\mu \nu}  \left(q,\xi  \right) \approx 2 \frac{\not \! k}{\left((k+q)^2+i \epsilon\right)\left(k^2+i\epsilon\right) } k^\nu
G_{\mu \nu}  \left(q,\xi  \right) \approx 0, \label{gapprox}
\ee
because $k^\nu \approx k^+ \xi^\nu$ with corrections proportional to $\lambda p^+, \lambda \ll 1$, as follows from the power counting described in Sect. \ref{powercount}, and because $ G_{\mu \nu}  \left(q,\xi  \right) \xi^\nu = 0$. An analogous observation holds for the coupling of $G$-gluons to jet-lines via triple-gluon vertices. In (\ref{gapprox}) we neglect all terms of order $\lambda p^+$ in the numerator, including the momentum $q$, because we assume that the denominator also scales as $\sim \lambda p^+$. This approximation is only valid for soft gluons not in the Glauber or Coulomb region, where the denominator behaves as $\sim \lambda^2 p^+$.
The remaining step in order to obtain a fully factorized form of the PDFs is therefore to show that the soft momenta are not pinched in this Glauber/Coulomb region. Then we can deform the integration contours over these momenta away from this dangerous region, into a purely soft region where the above approximations are applicable.

Consider again the 3-point vertex in Eq. (\ref{gapprox}), where now the gluon with momentum $q$ is in the Glauber/Coulomb region. If $|k^+ q^-|$ is not dominant over $|2 k_\perp \cdot q_\perp + q_\perp^2|$ in the denominator our approximation fails. The poles of the participating denominators are in the $q^-$ complex plane at
\ba
q^- & = & \frac{q_\perp^2 - i \epsilon}{2 q^+}, \nonumber \\
q^- & = & \frac{(k_\perp+q_\perp)^2 - i \epsilon}{2 (k^+ + q^+)} - k^- .
\ea
As long as the jet-line $k$ carries positive plus momentum, we see that the $q^-$-poles are not pinched in the Glauber region. In this case we can deform the contour away from this region into the purely soft region, where $ |k^+ q^-| \gg |2 k_\perp \cdot q_\perp + q_\perp^2|$. In a reduced diagram, which represents a physical process, there must be at every vertex at least one line which flows into the vertex, and at least one line which flows out of the vertex. Thus, at every vertex,  we can always find a momentum $k$ for which the above observation holds, and we can always choose the flow of $q$ through the jet to the hard scattering along such lines. Because the soft gluon momenta $\{q_{L,\,m_L}\}$, which we want to decouple, connect only to the purely virtual, initial-state jet $J_p$\footnote{Initial and final states are defined with respect to the hard scattering.}, this observation remains true throughout the jet.  An analogous argument applies to the right of the cut.

In summary, soft gluons can be decoupled from initial state jets at leading power in $1-x$. We therefore arrive at the factorized form of the parton distribution function as $x \rightarrow 1$:
\ba
f_{f}^{x\rightarrow 1}\;(x) &\! = & \! H_L(p,\mu;\beta,\xi)\; H_R(p,\mu;\beta,\xi)  \bar{J}_{p,\,L}(p,\mu;\beta; \xi ) \; \bar{J}_{p,\,R}(p,\mu;\beta; \xi)  \nonumber \\
& \times & \;  \sum\limits_{C_\beta,C_S}
\prod\limits_{m_L,m_R,l} \int \frac{d^n q_{L,\,m_L}}{(2 \pi)^n} \frac{d^n q_{R,\,m_R}}{(2 \pi)^n} \frac{d^n \bar{q}_l}{(2 \pi)^n} \;
\mathcal{E} \left(\xi, \{q_{L,\,m_L}^-\} \right)^{\{\gamma_{m_L}\}} \mathcal{E}^{*} \left(\xi, \{q_{R,\,m_R}^-\} \right)^{\{\gamma_{m_R}\}}
\nonumber \\
& \times & \; \int dy \int dz \; S^{(C_S)} \left( y p,\mu; \{q_{L,\,{m_L}}^{\gamma_{m_L}} \}; \{q_{R,\,{m_R}}^{\gamma_{m_R}} \};\{\bar{q}_l^{\delta_l}\} \right) \; \tilde{J}^{(C_\beta)}_{\beta}(z p, \mu;\xi;\{\bar{q}_l^{\delta_l}\}) \nonumber
  \\
& \times &  \; \delta^n\left(\sum_{m_L} q^\mu_{L,\,m_L} + \sum_{m_R} q^\mu_{R,\,m_R} +  \sum_l \bar{q}^\mu_{l} \right) \; \delta(1-x-y-z) \;,\label{factform2}
\ea
where we have grouped the eikonal factors stemming from the soft approximation with the soft function and the eikonal jet. We define
\be
\tilde{J}_{p,\,L}(p,\mu;\beta; \{q_{L,\,{m_L}}^{\gamma_{m_L}} \} ) =  \mathcal{E} \left(\xi, \{q_{L,\,m_L}^-\} \right)^{\{\gamma_{m_L}\}} \;\bar{J}_{p,\,L}(p,\mu;\beta; \xi ),  \label{bardef}
\ee
and analogously for the jet to the right of the cut, $J_{p,\,R}$, with a complex conjugate eikonal.

\subsubsection{Fully Factorized Form with an Eikonal Cross Section}

Although in Eq. (\ref{factform2}) the various functions are clearly separated in their momentum dependence, the parton distribution function is not quite in the desired form yet. We want to write the PDF in terms of a color singlet eikonal cross section, built from ordered exponentials:
\ba
\sigma^{(\mbox{\tiny eik})}_{a a}\left(\frac{(1-x)p^+}{\mu},\alpha_s(\mu),\varepsilon\right) & = & \frac{p^+}{\mbox{Tr } \mathbf{1}} \int \frac{d y^-}{2 \pi} e^{i(1-x)p^+ y^-}\nonumber \\
& & \qquad \times  \mbox{ Tr } \left< 0 \left| \bar{\mbox{T}} \left[ \mathcal{W}^{(aa)} (0,y^-,0_\perp)^\dagger \right] \; \mbox{T} \left[ \mathcal{W}^{(a a)} (0) \right] \right| 0 \right>, \label{eiksigdef}
\ea
where the product of two non-Abelian phase operators (Wilson lines) in the representation $a$, for quarks, is defined as follows:
\ba
\mathcal{W}^{(aa)}(x) & = &   \Phi^{(a)}_\beta (\infty,0;x)\; \Phi^{(a)}_\xi (0,-\infty;x),  \\
\Phi^{(f)}_\beta (\lambda_2,\lambda_1;x) & = & P e^{-i g \int_{\lambda_1}^{\lambda_2} d \lambda \beta \cdot {\mathcal{A}^{(f)}} (\lambda \beta + x )},
\ea
where the light-like velocities $\xi$ and $\beta$ are defined in (\ref{lightdef}), and where ${\mathcal{A}}^{(f)}$ is the vector potential in the representation of a parton with flavor $f$.  The trace  in (\ref{eiksigdef}) is over color indices. The lowest order of the eikonal cross section is normalized to $\delta(1-x)$.
This eikonal cross section has ultraviolet divergences which have to be renormalized, as indicated by the renormalization scale $\mu$. Furthermore, the delta-function for the soft momenta in Eq. (\ref{factform2}) constrains the momentum of the final state in $\sigma^{(\mbox{\tiny eik})}$ to be $(1-x)p^+$.

We can factorize the eikonal cross section (\ref{eiksigdef}) in a manner analogous to the full parton distribution function, and obtain
\ba
\sigma^{(\mbox{\tiny eik})}_{aa}\left(1-x\right) & = & \sum\limits_{C_\beta,C_S}
\prod\limits_{m_L,m_R,l} \int \frac{d^n q_{L,\,m_L}}{(2 \pi)^n} \frac{d^n q_{R,\,m_R}}{(2 \pi)^n} \frac{d^n \bar{q}_l}{(2 \pi)^n} \; \nonumber \\
& \times & \;
\tilde{J}^{(\mbox{\tiny eik})}_{p,\,L}(\xi,\mu;\beta; \{q_{L,\,{m_L}}^{\gamma_{m_L}} \} )
 \; \tilde{J}^{(\mbox{\tiny eik})}_{p,\,R}(\xi,\mu;\beta; \{q_{R,\,{m_R}}^{\gamma_{m_R}} \} )\; \nonumber \\
& \times & \; \int dy \int dz\; S^{(C_S)} \left(yp,\mu; \{q_{L,\,{m_L}}^{\gamma_{m_L}} \}; \{q_{R,\,{m_R}}^{\gamma_{m_R}} \};\{\bar{q}_l^{\delta_l}\} \right)
\;
\tilde{J}^{(C_\beta)}_{\beta}(zp,\mu;\xi;
\{\bar{q}_l^{\delta_l}\})
\;  \nonumber
  \\
& \times & \;
\delta^n\left(\sum_{m_L} q^\mu_{L,\,m_L} + \sum_{m_R} q^\mu_{R,\,m_R} +  \sum_l \bar{q}^\mu_{l} \right) \; \delta(1-x-y-z).  \label{eikfact}
\ea
The eikonal jets $\tilde{J}^{(\mbox{\tiny eik})}_{p,\,L}$, $\tilde{J}^{(\mbox{\tiny eik})}_{p,\,R}$ moving collinear to the momentum $p$, are defined analogously to Eq. (\ref{Jp}), with the fermion line carrying momentum $p$ replaced by an eikonal line in representation $a$ with velocity $\beta$. We can define analogous to Eq. (\ref{bardef})
\be
\tilde{J}^{(\mbox{\tiny eik})}_{p,\,L}(\xi,\mu;\beta; \{q_{L,\,{m_L}}^{\gamma_{m_L}} \} ) =  \mathcal{E} \left(\xi, \{q_{L,\,m_L}^-\} \right)^{\{\gamma_{m_L}\}} \;\bar{J}^{(\mbox{\tiny eik})}_{p,\,L}(\mu;\beta; \xi ),  \label{bardefeik}
\ee
and similarly for the jet to the right of the cut, $J_{p,\,R}^{(\mbox{\tiny eik})}$, with a complex conjugate eikonal. In the following we suppress the index $a$ for better readability.

Combining Eqs. (\ref{factform2}), (\ref{eikfact}), and (\ref{bardefeik}),  we arrive at the final form of the factorized parton distribution function, shown in Fig. \ref{partonfactend},
\ba
f_{f}^{x\rightarrow 1}\;(x) & = &  H_L\left( p, \mu \right)\; H_R\left(p, \mu \right) \; J_{p,\,L}^R (p,\mu)\; J_{p,\,R}^R (p,\mu)\; \sigma^{(\mbox{\tiny eik})} \left((1-x)p,\mu\right), \label{finalform}
\ea
suppressing the dependence on the lightlike vectors (\ref{lightdef}).
The purely virtual jet-remainders are defined by
\be
J_{p,\,L}^R (p,\mu) = \frac{\bar{J}_{p,\,L}(p,\mu;\beta; \xi )  }
{ \bar{J}^{(\mbox{\tiny eik})}_{p,\,L}(\mu;\beta; \xi )}\;. \label{jetvirdef}
\ee

\begin{figure}
\begin{center}
\epsfig{file=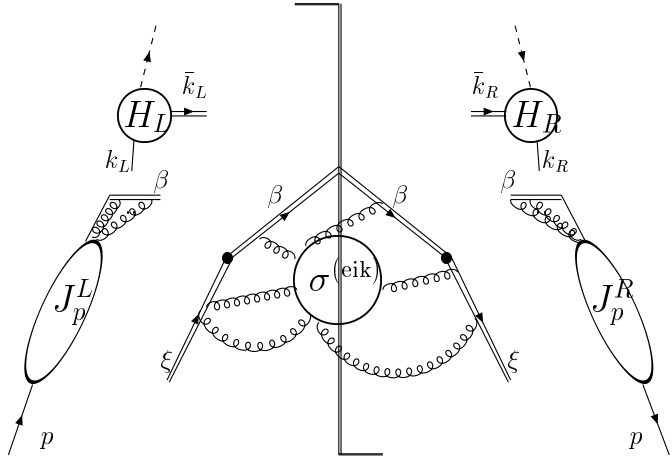,height=7.0cm,clip=0}
\caption{Parton distribution function for $x \rightarrow 1$, factorized into hard scatterings, an eikonal cross section, and purely virtual jet-remainders, as derived in Eq. (\ref{finalform}). The virtual jet functions are normalized by their eikonal analogs, as in Eq. (\ref{jetvirdef}).} \label{partonfactend}
\end{center}
\end{figure}

The contributions to the various functions in Eq. (\ref{finalform}) can be extracted from any Feynman diagram by examining each region of integration separately. In the following, we only need the eikonal cross section in the form (\ref{eiksigdef}), as well as the $x$-dependence and renormalization properties of the PDF and the terms constituting its factorized form. These properties are the subject of the next subsection.

\subsection{Renormalization of Parton Distribution Functions}

As was shown in \cite{pdfs}, the parton distribution functions,  defined in their unrenormalized form in  terms of nonlocal operators (\ref{pdfdef}), obey the evolution  equation (\ref{evol}), where the kernel $P_{ab}$ is found from the usual relation \cite{pdfs,bookTASI}
\be
P_{ff}(\alpha_s,x) = A_f(\alpha_s) (x)  \left[ \frac{1}{1-x} \right]_+  +  \dots =  - \frac{1}{2} g \frac{\partial }{\partial g} \ln Z^A_1 \left[ \frac{1}{1-x} \right]_+ + \dots, \label{prela}
\ee
where $g^2/(4 \pi) = \alpha_s$, and where $\ln\, Z^A_1$ denotes the $\frac{1}{\varepsilon}$-pole of the counterterm which multiplies the plus-distribution, plus scheme dependent constants, if we work in a minimal subtraction scheme with dimensional regularization. Above we only exhibit the term which is singular as $x \rightarrow 1$, since it is this term which we want to extract from the renormalization of our factorized form, Eq. (\ref{finalform}).

From Eq. (\ref{finalform}) we observe that only the eikonal cross section can contribute to the $A$-term proportional to a plus-distribution. This is because the hard functions are off-shell by $\mathcal{O}\left(x p^+\right)$, and the jet-remainders are purely virtual, thus cannot contain plus-distributions. Therefore, their renormalization has to be proportional to $B_f \, \delta(1-x)$, as was observed in \cite{EGW}.

It is thus the renormalization of a color singlet eikonal vertex which we have to study, in order to compute the coefficients $A^{(n)}$ in (\ref{aaa}). Although there are, aside from the usual QCD divergences, additional divergences at the eikonal vertex, this is a significant simplification compared to the investigation of the UV behavior of the full PDF. Let us now study the eikonal cross section as defined in Eq. (\ref{eiksigdef}), whose perturbative expansion follows the Feynman rules shown in Fig. \ref{Frules}.

\section{Nonabelian Eikonal Exponentiation} \label{sectexp}

Although the eikonal approximation simplifies the perturbative calculation already significantly, there are still many diagrams to be calculated at each order. In addition, the eikonal approximation also introduces new infrared divergences.

An observation first made by Sterman \cite{George}, then proved by
Gatheral \cite{gath}, and Frenkel and Taylor \cite{freta} solves both problems. This theorem states that a cross section $X$ with two eikonal lines in a nonabelian theory exponentiates,
\begin{equation}
\sigma^{(\rm eik)} \equiv X = e^{Y}, \label{eq1}
\end{equation}
where $Y$ can be given a simple recursive definition.
In this section we will recall the proof of Eq. (\ref{eq1}) \cite{gath,freta}, for the sake of completeness, including a few illustrative examples which will be used below for the calculation of the 2-loop coefficient $A^{(2)}$. The exponent $Y$ in (\ref{eq1}) has the following properties:
\begin{enumerate}
\item $Y$ is a subset of the diagrams contributing to $X$, which we will call ``webs'' in the following, since, as we will see below, their lines
`` [...] are all nested [...] in a spider's web pattern'' \cite{George}.
\item The color weights of the diagrams in $Y$ are in general different from those in $X$.
\item For Eq. (\ref{eq1}) to hold, the phase-space region should be symmetric in the real gluon momenta.
\end{enumerate}

Below we will outline the arguments necessary to prove this theorem.  The proof relies on the recursive definition of color-weights and on the iterated application of a  well-known eikonal identity. Then we go on to show that IR and UV subdivergences cancel in the exponent at each order.

\subsection{Proof of Exponentiation}

\subsubsection{Some Terminology} \label{defs}

In order to specify which subset of diagrams of the original perturbation series $X$ contributes to the exponent $Y$ we need to introduce some terminology.

Each diagram will be decomposed into its color part and its Feynman integral in the eikonal approximation. The eikonal Feynman rules were given above in Fig. \ref{Frules}. The color part can be represented graphically in a diagram which is similar to an ordinary Feynman diagram, but the vertices represent the color part of the Feynman rules, i.\,e. the vertices are just the $T^a_{ij}$s and $i\, f_{ijk}$s, for quark or gluon, respectively, and the lines are $\delta_{ij}$s. In addition, all soft lines have to be drawn inside the (cut) eikonal loop for reasons which will become clear shortly. Certain color diagrams are related to each other by use of the commutation relations of the $T^a_{ij}$s and $i\, f_{ijk}$s (Jacobi identity) which are graphically represented in Fig. \ref{commrel}.
\ba
\left[ T^a, T^b \right] & = & i f_{abc} T^c \nonumber \\
f_{ilm} f_{mjk} &+ & f_{jlm} f_{imk} + f_{klm} f_{ijm} = 0. \label{jacid}
\ea

\begin{figure}[htb]
\begin{center}
\epsfig{file=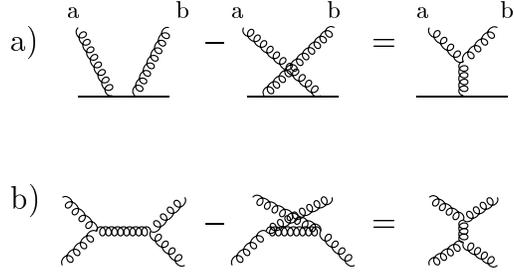,height=5cm,clip=0}
\caption{Graphical representations of the commutation relations between a) $T^a_{ij}$s, and b)  $i\, f_{ijk}$s (Jacobi identity), Eq. (\ref{jacid}).} \label{commrel}
\end{center}
\end{figure}

As mentioned above, the diagrams contributing to $Y$ will be called ``webs'' \cite{George}. Originally \cite{gath},  a web was defined as a set of gluon lines which cannot be partitioned without cutting at least one of its lines. As already stated above, all soft lines are to be drawn inside the eikonal loop(s). However, at ${\mathcal{O}}\left(\alpha_s^3\right)$ new types of diagrams arise which Frenkel and Taylor \cite{freta} called connected webs (``c-webs''). c-webs are not included in the original definition for the following reason: If one cuts the horizontal gluon line of the c-web drawn in Fig. \ref{webexample} one would get two webs consisting of three-point vertices since real and virtual gluon lines are treated on equal footing in a color-weight diagram. Below we will refer to webs and c-webs just as webs.

The definitions given by Gatheral, and Frenkel and Taylor can be unified by the following \emph{definition}:
\vspace*{1mm}

A web is a (sub)diagram consisting of soft gluon lines connecting two eikonal lines which cannot be partitioned into webs of lower order by cutting all eikonal lines exactly once. Stated differently, webs are two-eikonal irreducible diagrams. The \emph{order of a web} is defined to be equal to the powers  of $\alpha_s$ it contains, e.\,g. a web of ${\mathcal{O}}\left(\alpha_s^2\right)$ will be called a web of order 2. Diagrammatic examples are shown in Fig. \ref{webexample}. A web has a color factor $\overline{C}$ and a Feynman integral part $\mathcal{F}$. $\mathcal{F}$ contains only those eikonal propagators which are \emph{internal} to the web. The color weight is in general different from the one which one would get from the usual Feynman rules.
\vspace*{1mm}

\begin{figure}[hbt]
 \begin{center}
\epsfig{file=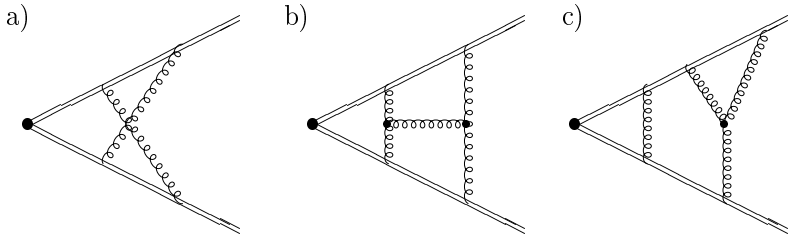,height=4cm,clip=0}
\caption{Examples of a) a web (order 2), b) a c-web (web of order 3), and c) a diagram which is not a web (consisting of a product of an order-1 web and of an order-2 web).} \label{webexample}
\end{center}
\end{figure}

The color weight of  a web of order $m$ is recursively defined as
\begin{eqnarray}
\overline{C}(W^{(m)}) & \equiv & \frac{1}{\mbox{Tr } \mathbf{1}} C(W^{(m)}) - \sum_d  \prod\limits_{n_i} \overline{C} (W_{n_i}^{(i)}), \nonumber  \\
\overline{C}(W^{(1)}) & \equiv & \frac{1}{\mbox{Tr } \mathbf{1}} C(W^{(1)}) ,
\label{colwe}
\end{eqnarray}
where $C(W^{(m)})$ is the ordinary color factor, $\frac{1}{\mbox{\tiny Tr } \mathbf{1}}$ is the usual normalization (see Eq. (\ref{eiksigdef})), $C^{(0)} = \mbox{Tr } \mathbf{1}$, $\sum_d$ is the sum over the set of all non-trivial decompositions $d$ of $W^{(m)}$ into webs of order $i < m$, and $\prod\limits_{n_i}$  denotes the product of all webs $n_i$ of order $i$ ($1 \leq i < m$) in a particular decomposition $d$. The set of all non-trivial \emph{decompositions} of a given web can be obtained by successively disentangling crossed gluon lines in the web by repeated application of the color identities given in  Fig. \ref{commrel}.

In \cite{gath} Gatheral showed that webs in the original definition have what he called ``maximally nonabelian'' color weights $\sim \alpha_s^m C_F C_A^{m-1}$, where the $C_i$s are the Casimir factors in the fundamental and adjoint representation, respectively. This statement, however, is misleading at orders  $> \alpha_s^3$ \cite{freta}. We will see an example below, in the calculation of the $N_f$ term contributing to the coefficient $A$ at three loops.

In the following subsection we will clarify  the above definitions in an example which shows how to factorize eikonal Feynman diagrams into sums of products of webs. This then leads directly to exponentiation. The recursive definition of the color weights of the webs ensures the factorization of the color parts. For the factorization of the Feynman eikonal integrals $\mathcal{F}$ we will make repeated use of the eikonal identity  \cite{levsuch}
\begin{equation}
 \frac{1}{p \cdot k_1} \frac{1}{p \cdot (k_1 + k_2)} +  \frac{1}{p \cdot k_2} \frac{1}{p \cdot (k_1 + k_2)} = \frac{1}{p \cdot k_1} \frac{1}{p \cdot k_2} \label{eikid},
\end{equation}
illustrated in Fig. \ref{rules} a). This identity can be extended to an arbitrary number of soft gluons in a straightforward way by repeated application of Eq. (\ref{eikid}): For two webs $W_1$ and $W_2$ with gluon legs $k_i \,(i = 1,\dots,m)$ and $l_j\,(j=1,\dots, n)$ attached to an eikonal line with velocity $p$ the generalized identity reads
\begin{eqnarray}
{{\mathcal{F}}}(W_1) {{\mathcal{F}}}(W_2) & \sim  &  \frac{1}{p \cdot k_1 \, p \cdot (k_1 + k_2) \dots p \cdot (k_1 + \dots + k_m)}  \frac{1}{p \cdot l_1 \, p \cdot (l_1 + l_2) \dots p \cdot (l_1 + \dots + l_n)} \nonumber \\
& = & \sum\limits_{\mbox{\tiny perms} (n,m)} F, \label{gen2eikid}
\end{eqnarray}
where the sum is over all Feynman diagrams $F$ obtained by permuting the $n+m$ gluon lines such that the order of the $k_i$, and $l_j$, respectively, \emph{within} each web is not changed. A simple example is shown in Fig. \ref{rules} b). The extension to more than two webs follows by repeating the above argument:
\begin{equation}
\sum\limits_{F \,  \mbox{\tiny in } d} F = \prod\limits_{n_i} {\mathcal{F}} \left( W_{n_{i}}^{(i)} \right). \label{geneikid}
\end{equation}

\begin{figure}[htb]
\begin{center}
\epsfig{file=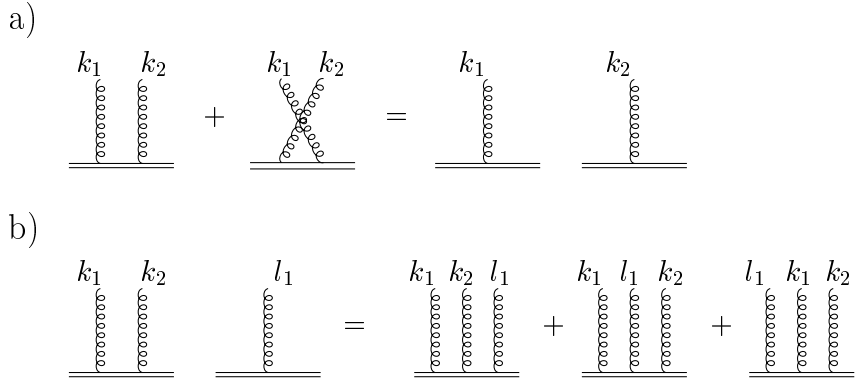,height=5cm,clip=0}
\caption{a) Eikonal identity for 2 gluons, and b) an illustration of the generalized eikonal identity Eq. (\ref{gen2eikid})  for two webs. } \label{rules}
\end{center}
\end{figure}

\subsubsection{An Example}    \label{examplesubsect}

We will show by induction that the terms in the perturbation series $X$ in Eq. (\ref{eq1}), normalized by the zeroth order contribution, can be reorganized into a sum of products of webs which can be rewritten as $\exp(Y)$. Therefore it is necessary and also instructive to start with the first nontrivial example, diagrams of ${\mathcal{O}}(\alpha_s^2)$. At this order we have the Feynman diagrams shown in Fig. \ref{order2}, for quark or antiquark eikonal lines, excluding eikonal and gluon self-energies. Eikonal self-energies vanish if we work in Feynman gauge. The sum of diagrams at a given order in $\alpha_s$ is gauge invariant, of course. The contribution of the first two terms in Fig. \ref{order2} can be rearranged by applying Eqs. (\ref{colwe}) and (\ref{eikid}) as shown in Fig. \ref{rearr}.

\begin{figure}
\begin{center}
\epsfig{file=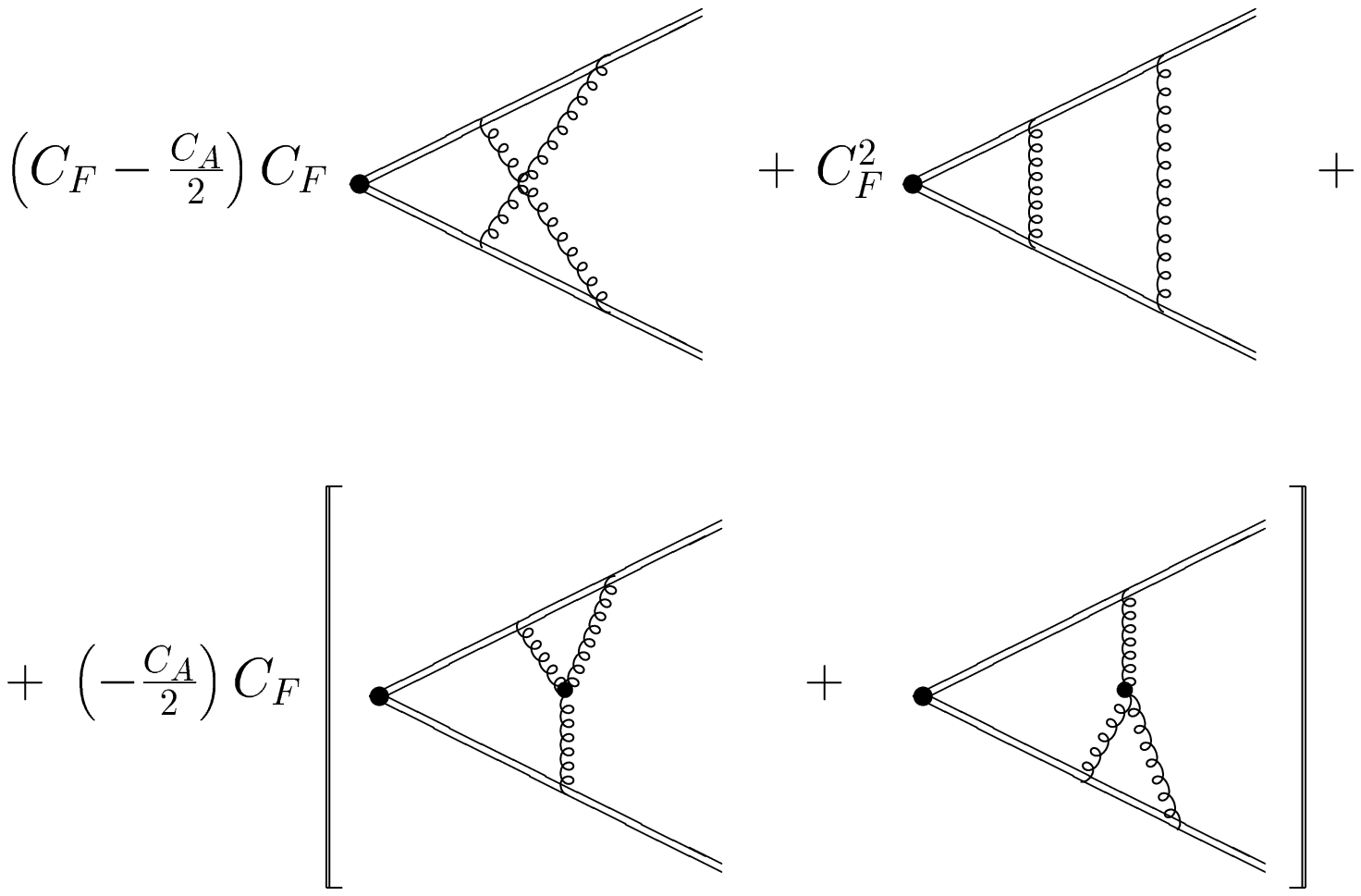,height=6.3cm,clip=0}
\caption{Diagrams contributing at ${\mathcal{O}}(\alpha_s^2)$ (excluding self-energies). The color factors are given for eikonal lines in the fundamental representation, omitting the overall normalization $\frac{1}{\mbox{Tr } \mathbf{1}}$ everywhere.} \label{order2}
\end{center}
\end{figure}

\begin{figure}
\begin{center}
\epsfig{file=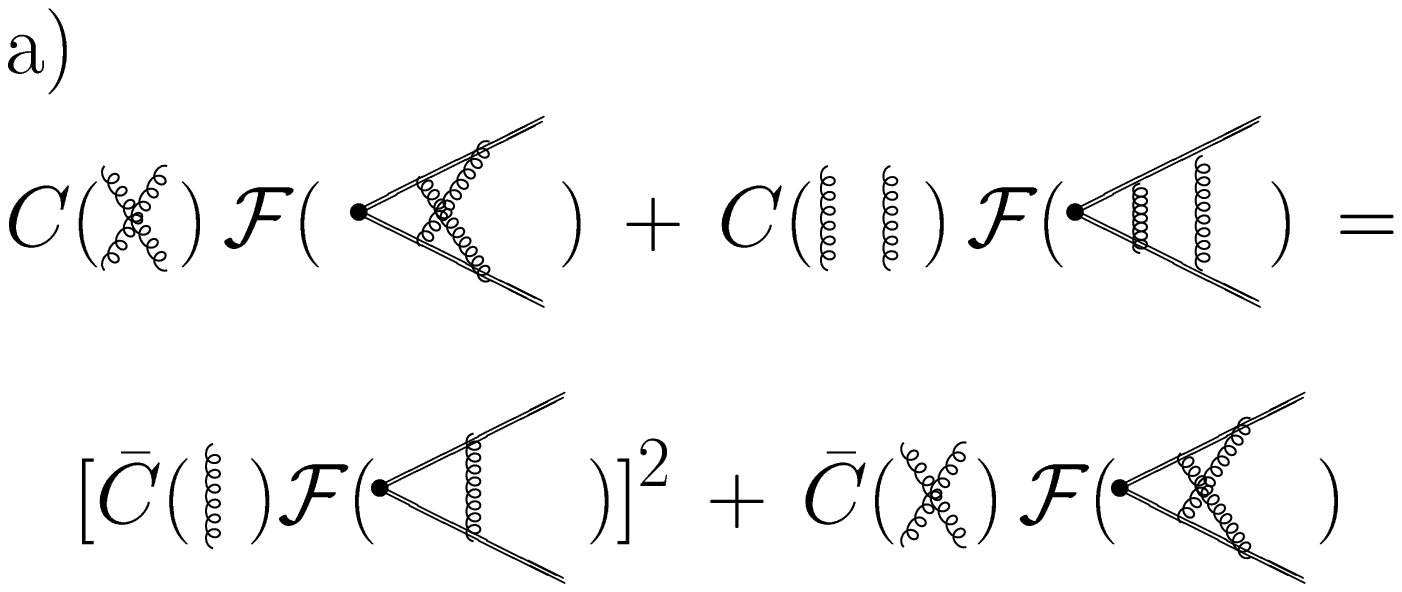,height=3.6cm,clip=0}
\\
\hspace*{-9mm} \epsfig{file=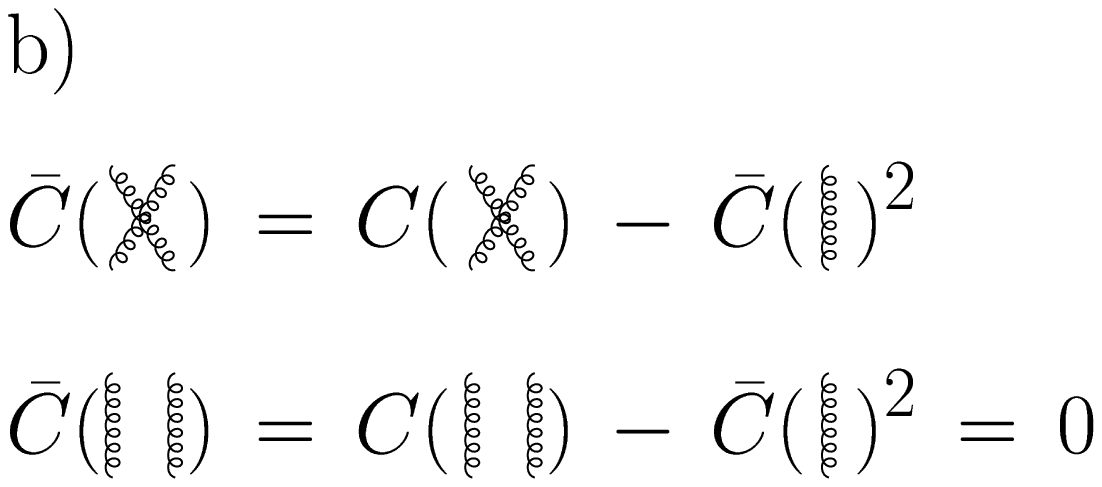,height=3.3cm,clip=0}
\caption{a) Rearrangement of the first two terms of Fig. \ref{order2} using Eq. (\ref{eikid}), and b) Eq. (\ref{colwe}).} \label{rearr}
\end{center}
\end{figure}

Thus we arrive at the following series obtained by rearranging the expansion of $X$ up to ${\mathcal{O}}(\alpha_s^2)$:
\begin{eqnarray}
X & = & \mathbf{1} + \sum\limits_{\stackrel{\mbox{\tiny all webs }}{\mbox{\tiny of order } 1}}  \overline{C} \left(W^{(1)} \right) {\mathcal{F}} \left(W^{(1)} \right) + \nonumber \\
& + & \frac{1}{2 !} \bigg(  \sum\limits_{\stackrel{\mbox{\tiny all webs }}{\mbox{\tiny of order } 1}}  \overline{C} \left(W^{(1)} \right) {\mathcal{F}} \left(W^{(1)} \right) \bigg)^2 +  \sum\limits_{\stackrel{\mbox{\tiny all webs }}{\mbox{\tiny of order } 2}}  \overline{C} \left(W^{(2)} \right) {\mathcal{F}} \left(W^{(2)} \right) + \dots, \label{exporder2}
\end{eqnarray}
which is illustrated in Fig. \ref{expgraph}. The combinatorical factor $\frac{1}{2!}$ is necessary to avoid overcounting since two webs with the same structure are indistinguishable if the integration measure is symmetric in the real gluon momenta.

\begin{figure}[htb]
\begin{center}
\epsfig{file=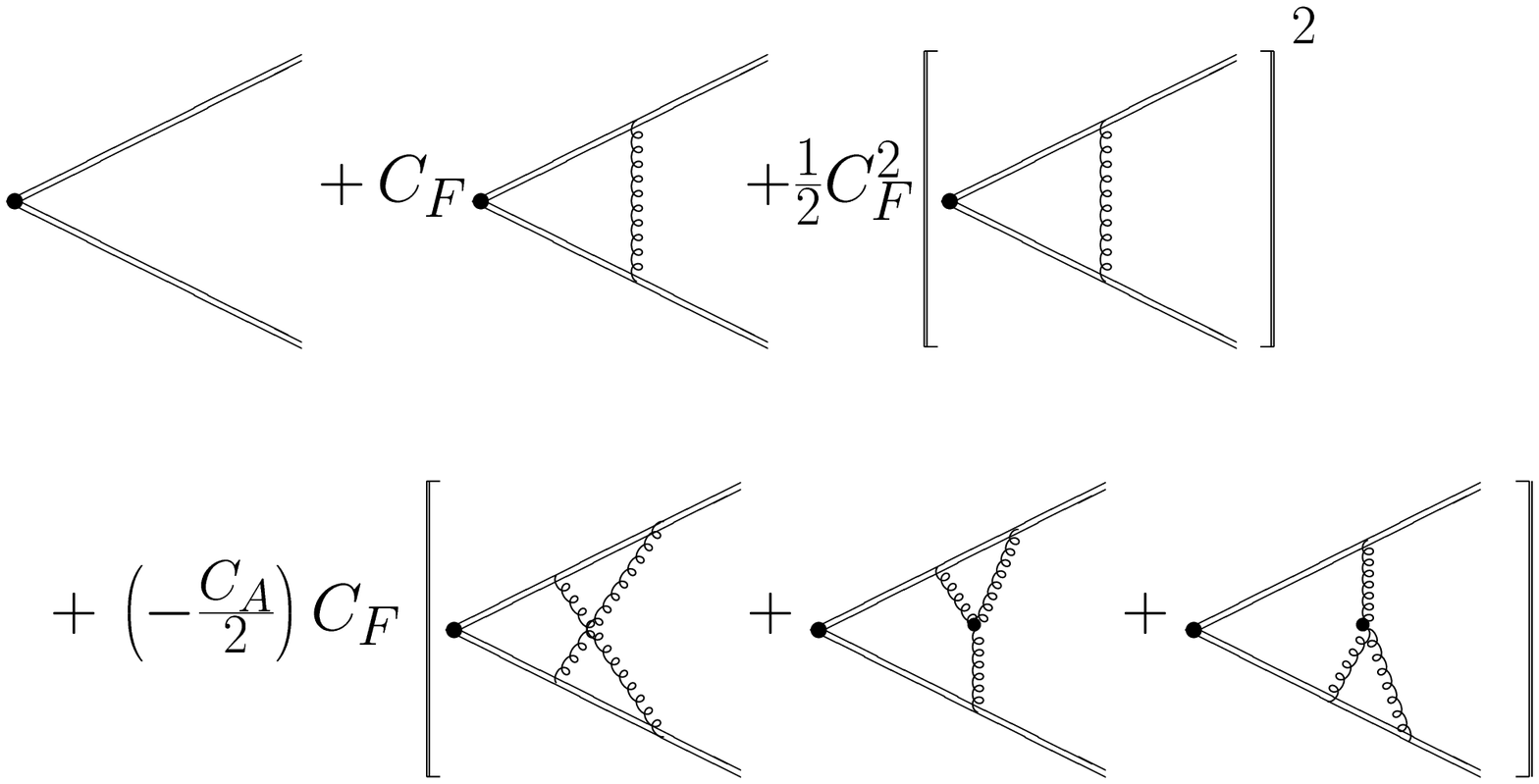,height=5.4cm,clip=0}
\caption{Graphical expression of the exponentiation up to webs of order 2 in Feynman gauge. The color weights are given for (anti)quark eikonal lines. } \label{expgraph}
\end{center}
\end{figure}


\subsubsection{Exponentiation}

Looking at the above example it is now clear that any Feynman diagram with two eikonal lines can be expressed as a sum of products of webs by applying Eqs. (\ref{colwe}) and (\ref{geneikid}) repeatedly. By induction we arrive at the following equation for the set of  Feynman diagrams of ${\mathcal{O}}(\alpha_s^n)$, $F^{(n)}$:
\begin{equation}
F^{(n)} = \sum\limits_{\left\{ n_i \right\} } \prod_i \frac{1}{n_i !} \bigg( \sum\limits_{\stackrel{\mbox{\tiny all webs}}{\mbox{\tiny of order }i}} \overline{C}(W^{(i)}) {\mathcal{F}}(W^{(i)}) \bigg)^{n_i}, \label{nfac}
\end{equation}
where $i$ labels the order of the webs and the sum is over all sets $\left\{ n_i \right\},\, 0 \leq n_i < \infty$ such that $\sum_i i \, n_i = n$. For example, at ${\mathcal{O}} (\alpha_s^3)$ we can have $n_1 = 3$ webs of order 1 ($\{3,0,\dots\}$), or $n_1 = 1$ webs of order 1 and $n_2 = 1$ webs of order 2 ($\{1,1,0,\dots\}$), or $n_3 = 1$ webs of order 3 ($\{0,0,1,0,\dots\}$). The combinatorical factor of $\frac{1}{n!}$ is needed to avoid overcounting because of  property 3.) of $X$,  stated in the introduction to this section, namely that the integration measure is symmetric in the real gluon momenta, for example ${\mathcal{F}}^{(1)}(k_1) {\mathcal{F}}^{(2)}(k_2, k_3) = {\mathcal{F}}^{(1)}(k_2) {\mathcal{F}}^{(2)}(k_1, k_3)$, which means that webs of the same structure are indistinguishable. Were property 3.) not fulfilled, the perturbation series would not exponentiate.

We now rearrange the original perturbation series given in powers of $\alpha_s^n$
\begin{equation}
X = \sum\limits_{n=0}^\infty F^{(n)}, \label{ser}
\end{equation}
\begin{eqnarray}
X & = & \sum\limits_{n=0}^\infty \sum\limits_{\left\{ n_i \right\} } \delta_{n \sum_i i n_i}
\prod_i \frac{1}{n_i !} \bigg( \sum\limits_{\stackrel{\mbox{\tiny all webs}}{\mbox{\tiny of order }i}} \overline{C}(W^{(i)}) {\mathcal{F}}(W^{(i)}) \bigg)^{n_i} \nonumber \\
& = & \sum\limits_{\stackrel{\mbox{\tiny all possible }}{ \{n_i\}}}
\prod_i \frac{1}{n_i !} \bigg( \sum\limits_{\stackrel{\mbox{\tiny all webs}}{\mbox{\tiny of order }i}} \overline{C}(W^{(i)}) {\mathcal{F}}(W^{(i)}) \bigg)^{n_i}    \nonumber
\\
& = & \prod_i \left\{ \sum_{n_i} \frac{1}{n_i !} \bigg( \sum\limits_{\stackrel{\mbox{\tiny all webs}}{\mbox{\tiny of order }i}} \overline{C}(W^{(i)}) {\mathcal{F}}(W^{(i)}) \bigg)^{n_i} \right\} \nonumber \\
&  = & \prod\limits_{i} \exp \bigg( \sum\limits_{\stackrel{\mbox{\tiny all webs}}{\mbox{\tiny of order }i}} \overline{C}(W^{(i)}) {\mathcal{F}}(W^{(i)}) \bigg),
\end{eqnarray}
where we have used the fact that for any function $f(n_i,i)$
\be
\sum\limits_{\stackrel{\mbox{\tiny all possible }}{ \{n_i\}}}  \prod_i f(n_i,i) = \prod_i \sum_{n_i} f(n_i,i)
\ee
which is easy to see by comparing the expansions of the left and the right hand sides.

So the series exponentiates
\begin{equation}
X = e^Y, \quad Y \equiv \sum\limits_{i} \bigg( \sum\limits_{\stackrel{\mbox{\tiny all webs}}{\mbox{\tiny of order }i}} \overline{C}(W^{(i)}) {\mathcal{F}}(W^{(i)}) \bigg). \label{exp2eq}
\end{equation}
This completes the proof that eikonal cross sections with two eikonal lines can be written as an exponent of an infinite sum of webs.

\subsection{Cancellation of Subdivergences in the Exponent}  \label{wardidproof}

Gatheral, Frenkel, and Taylor showed in  \cite{gafreta} by explicit fixed-order calculations that
infrared/ collinear subdivergences cancel in the exponent. Here we will outline the proof of this
cancellation, as well as of the cancellation of UV subdivergences involving the eikonal vertex,
to all orders with the help of the identities in Fig. \ref{wardeik} in the soft approximation.
The remaining UV subdivergences are removed via ordinary QCD counterterms, and thus an additional
investigation of the renormalizability of the eikonal vertex is unnecessary.

 To show the absence of subdivergences, let us rewrite
 Eq. (\ref{nfac}) as
\be
\sum\limits_{\mbox{\tiny order } n} \bar{C}^{(n)} \mathcal{F}^{(n)} = F^{(n)}_{\mbox{\tiny conv}} + F^{(n)}_{\mbox{\tiny div}}
- \sum\limits_{\left\{ n_i \right\},\, i < n } \prod_i \frac{1}{n_i !} \bigg( \sum\limits_{\stackrel{\mbox{\tiny all webs}}{\mbox{\tiny of order }i < n}} \overline{C}(W^{(i)}) {\mathcal{F}}(W^{(i)}) \bigg)^{n_i}. \label{eqsubfact}
\ee
Eq. (\ref{eqsubfact}) means, that the sum of all webs at order $n$ are given by the original perturbation series at that
 order where all lower-order webs have been subtracted out. The original perturbation series can be classified
 into terms without subdivergences, denoted by $F^{(n)}_{\mbox{\tiny conv}}$, and terms which contain
 subdivergences, $F^{(n)}_{\mbox{\tiny div}}$.

\begin{figure}[hbt]
\begin{center}
\epsfig{file=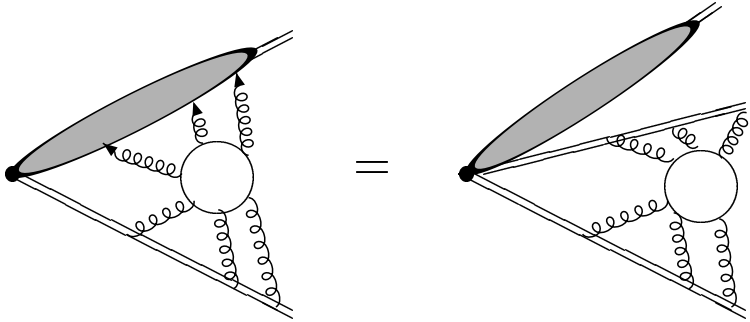,height=4.2cm,clip=0}
\caption{Factorization of jet-like collinear configurations, represented by the grey oval, from the eikonal cross section with the help of the soft approximation. } \label{subfact}
\end{center}
\end{figure}

In eikonal cross sections, infrared/collinear divergences stem from the same momentum configuration as UV divergences.
Since eikonal cross sections are scaleless, when a line becomes collinear to an eikonal it can carry infinite
momentum in a light-like direction.
 But in this case we can employ the soft approximation described in Sect. \ref{softapprox} to factorize
 these jet-like configurations from the rest of the eikonal cross section. The reasoning follows Section
 \ref{softapprox}, and we arrive at the equality shown in Fig. \ref{subfact}. The grey oval in the figure
 represents a specific jet-like configuration, collinear to one of the eikonal lines.  The displayed equality
 states that the sum of all webs at a given order, where this jet-like configuration is connected to the rest
 of the eikonal cross section by soft gluons, can be expressed in the factored form shown on the right-hand side.
 As in Section \ref{softapprox}, remainders  are non-leading. Due to the definition of the color weights
 (\ref{colwe}), the right-hand side does not constitute a web of the same order, but rather a product of webs
 of lower orders. The contribution shown on the left-hand side of Fig. \ref{subfact} is a contribution to
 $F^{(n)}_{\mbox{\tiny div}}$ of Eq. (\ref{eqsubfact}). In (\ref{eqsubfact}), however, we subtract out all products of webs of lower orders, thus cancelling the divergent contributions because of the equality shown in Fig. \ref{subfact}.
Using the equality in Fig. \ref{subfact} and Eq. (\ref{eqsubfact}) recursively for every IR/collinear
subdivergence, we see that the sum of webs at a given order is free of such subdivergences.

To summarize,
 the collinear configuration does not contribute at the order under consideration after summing over all relevant webs at that order, because this collinear configuration has already been taken into account at a lower order. The only possible collinear and UV vertex divergence can occur in the final, overall integral. Of course, in the original perturbative expansion $X$ of the eikonal cross section in Eq. (\ref{ser}) these collinear and UV subdivergent configurations contribute, but in the exponent $Y$ of Eq. (\ref{exp2eq}) they only appear as overall divergences.

\section{The Method for Obtaining $A^{(n)}_f$}    \label{sectmeth}

\subsection{Renormalization of Webs}

As was observed in the above sections, webs have at most one collinear/IR divergence, and one overall UV divergence from integration over the final momentum. Using the invariance of eikonal cross sections under rescalings of the eikonal velocities, and the fact that there is at most one overall IR divergence, we can deduce the dependence of the webs on its overall momentum $k$, and on the light-like eikonal momenta. For a color singlet eikonal cross section we have
\begin{equation}
W_{aa} \left( k^2, \frac{(k\cdot \beta_1) (k \cdot \beta_2)}{\beta_1 \cdot \beta_2}, \mu^2, \alpha_s(\mu^2), \varepsilon \right) =
 W_{aa} \left(k^2,k^2 + k_\perp^2,  \alpha_s(\mu^2),\varepsilon \right),
\end{equation}
after integration over internal momenta, in a frame where the eikonal lines are back-to-back \cite{EGW}. For definiteness, we pick the incoming line  $\xi$ moving in the plus direction, and the outgoing eikonal $\beta$ in the minus direction, and since quantities built from eikonal lines are scaleless, we can scale the eikonal velocities to 1. Recall (\ref{lightdef}):
\ba
\xi & = & \left( 1,0,0_\perp\right) \nonumber \\
\beta & = & \left( 0,1,0_\perp \right) \label{betaframe}
\ea
in light-cone coordinates. This choice will simplify the calculations considerably, as we will see below. We can therefore write the contributions from virtual webs of order $n$ to the eikonal cross section as
\ba
 2 \int \frac{d^{2 - 2\varepsilon} k_\perp}{(2 \pi)^{1-2 \varepsilon}} \int\limits_0^\infty \frac{d k^+}{2 k^+} \!\! & \int dk^2 \!\!& \!\!
 W_{aa}^{(n)} \left(k^2,k^2 + k_\perp^2,  \alpha_s(\mu^2),\varepsilon \right) \nonumber \\
& = & 2\; \bar{C}_a^{(n)} \left( \frac{\alpha_s(\mu^2)}{\pi} \right)^n \left(\mu^2\right)^{l \varepsilon} \left(4 \pi\right)^{l \varepsilon} K(\varepsilon)  \int \frac{d^{2 - 2\varepsilon} k_\perp}{(k_\perp^2)^{1+(l-1)\varepsilon}} \int\limits_0^\infty \frac{d k^+}{k^+},   \label{methodeq}
\ea
where $l$ is an integer $\leq n$, and $K$ contains numerical factors (including factors of $\pi$) and is, in general, a function of $\varepsilon$ due to the regulation of infrared and UV (sub)divergences.  Above, on the left hand side, all internal momenta have been integrated over, as well as $k^-$,  and internal UV divergences have been renormalized. The integration over $k^2$ results in terms $\sim \frac{1}{(k_\perp^2)^{(l-1)\varepsilon}}$. For graphs including (local) counterterms $l < n$, whereas for graphs with $n$ loops $l = n$. Both virtual webs and their complex conjugates contribute to the overall factor of 2. The structure of the integral over $k^+$ follows from boost invariance.  In Eq. (\ref{methodeq}), this integral is divergent, but these divergences cancel against the corresponding real contributions, and therefore do not affect the anomalous dimension of the eikonal vertex. The $k^+$-integral plays the role of $\frac{dx}{1-x}$ for the full  parton-in-parton distribution functions (cf. Eq. (\ref{eiksigdef})).

The final scaleless $k_\perp$ integral provides the $n$-loop UV counterterm which contributes to the anomalous dimension $P_{ff}$, Eq. (\ref{pff}). To isolate the UV pole we temporarily introduce a mass
\be
\int \frac{d^{2 - 2 \varepsilon} k_\perp}{(k_\perp^2+m^2)^{1+(l-1)\varepsilon}} = \pi^{1-\varepsilon} \frac{\Gamma(l \varepsilon)}{\Gamma(1 + (l-1)\varepsilon)} \left(m^2 \right)^{-l \varepsilon}. \label{scalepart}
\ee
The counterterm is then given, as usual, by minus the pole terms after expanding in $\varepsilon$. After summing over the contributions of all webs at a given order and their counterterms for subdivergences, all nonlocal terms ($\sim \ln \frac{\mu^2}{m^2}$) cancel as well as UV vertex counterterms and IR divergences, and we obtain the $n$-loop counterterm contributing at $x \rightarrow 1$, which can be written as a series in $\varepsilon$:
\be
Z^{(n)\,A} = \sum\limits_{m = 1}^n \frac{1}{\varepsilon^m} \left( \frac{\alpha_s(\mu^2)}{\pi} \right)^n a_m^{(n)} \int\limits_0^\infty \frac{d k^+}{k^+},
\ee
with purely numerical coefficients $a_m^{(n)}$. Because webs exponentiate, the counterterm for UV divergences in the perturbative expansion of a non-singlet parton distribution  is given by
\be
Z^A = \exp \left\{ \sum_{m = 1}^\infty \sum_{n = 1}^m \frac{1}{\varepsilon^m} \left( \frac{\alpha_s(\mu^2)}{\pi} \right)^n a_m^{(n)} \, \int\limits_0^\infty \frac{d k^+}{k^+}  \right\}  \label{counterterm}
\ee
in the limit $x \rightarrow 1$, as indicated by the superscript $A$.
Now it is trivial to extract the contribution to $P_{ff}$. From (\ref{prela}) and (\ref{counterterm}) we get
\be
A_f^{(n)} = - n a_1^{(n)}. \label{aendeq}
\ee

As emphasized above, internal UV divergences, including the usual QCD divergences and divergences at the eikonal vertex, have to be renormalized. Further complications arise because collinear/IR divergences cancel only after summing over all diagrams at a given order, so an individual diagram has in general UV singularities multiplying IR/collinear singularities. Our method to resolve these technical problems is most transparent in light-cone ordered perturbation theory (LCOPT) \cite{changma, kosop}, which is equivalent to performing all minus integrals of all loops. The Feynman rules for LCOPT are given in Appendix \ref{Lrules} for completeness.

\subsection{Summary of the Method in LCOPT}

Our method can be summarized as follows, details will be given below:
\begin{enumerate}
\item We start with the expressions in LCOPT for the set of webs at a given order with a fixed coupling. The number of web-diagrams is much less than the number of all possible diagrams at a given order. Moreover, since we work in Feynman gauge, the number of possible webs is further reduced. For example, at order 2, as we will see below, only three diagrams contribute, aside from gluon self-energies.
\item Ultraviolet divergent internal $k_{\perp,\,i}$-integrals are regularized via dimensional regularization, with $\varepsilon > 0$. At this stage we do not yet encounter IR/collinear singularities since all integrals over transverse momenta are performed at fixed plus momenta.
\item We add the necessary QCD counterterms and the counterterms for the eikonal vertex which has to be renormalized as a composite operator. As we showed in Section \ref{wardidproof}, the sum of the latter cancels because of the recursive definition of webs and a Ward identity. However, in the intermediate stages the vertex counterterms are necessary to make individual diagrams UV finite.
\item After elimination of the UV singularities we dimensionally continue to $\varepsilon < 0$ to regulate the IR/collinear plus-integrals. It follows from the rules for LCOPT that all internal plus momenta are bounded by the total $k^+$ flowing into the minus eikonal. Therefore, the integrals over these internal plus momenta give no UV subdivergences.
\item When we sum over the set of diagrams at a given order the IR divergent parts cancel, as well as the UV counterterms for the vertex, thus also the internally UV divergent vertex parts  cancel.
\item The final scaleless $k_\perp$ integral provides the UV counterterm contributing to the anomalous dimension (see Eqs. (\ref{methodeq})-(\ref{aendeq}) ).
\end{enumerate}

The rules for LCOPT can be found in Appendix \ref{Lrules} \cite{changma, kosop}, they can be derived by performing the minus integrals. LCOPT is similar to old-fashioned, or time-ordered, perturbation theory, but ordered along the light-cone, $x^+$, rather than in $x^0$. In a LCOPT diagram all internal lines are on-shell, in contrast to a covariant Feynman diagram, which allows us to identify UV divergent loops in eikonal diagrams more easily. A covariant Feynman diagram is comprised of one or more LCOPT diagrams.

We start by writing down all light-cone ordered diagrams of a given covariant Feynman diagram. All momenta in crossed gluon ladders have to be chosen independent of each other, such that they all flow through the eikonal vertex, since we seek the anomalous dimension for this vertex. Because $\xi$ has no minus-component (cf. Eq. (\ref{betaframe})), we have $q^- = 0$ in Eq. (\ref{denom}) of the appendix when applying the Feynman rules for LCOPT in our case. This can be depicted graphically by contracting all propagators on the minus-eikonal (here $\beta$) to a point, which coincides with the eikonal vertex. Two-loop examples can be found in Fig. \ref{2loopfig}. Sometimes, numerators stemming from triple-gluon vertices or quark propagators cancel the corresponding propagators on the plus-eikonal ($\xi \cdot k = k^-$). Graphically, this can again be described by contracting these propagators to a point. Then we can read off easily from the various light-cone ordered diagrams the analytical expressions, whose $k_{\perp,\,i}$-integrals we perform, and renormalize.

We need QCD counterterms and counterterms for the effective vertex. More specifically, by QCD counterterms we mean the usual gluon self-energy counterterms, as well as the counterterms for triple-gluon vertices and eikonal-gluon-eikonal vertices. The latter are UV divergent in any covariant diagram, however, this is not necessarily the case for all LCOPT diagrams found from a covariant diagram. Examples will be given below. Self-energies of the light-like eikonal lines vanish in Feynman gauge. Both types of counterterms are found via the (recursive) BPHZ-formalism, and the subdivergences are identified with the help of naive power-counting on a graph-by-graph basis.

We will now illustrate our method by the rederivation of the 1- and 2-loop $A$-coefficients.

\subsection{Calculation of the 1- and 2-loop Coefficients $A^{(1)}_f$, $A^{(2)}_f$} \label{sectexample}

The well-known \cite{2loopknown} coefficients of the collinear parts of the splitting functions to one and two-loop order are given by
\begin{eqnarray}
A_a^{(1)} & = & C_a,  \label{a1number} \\
A_a^{(2)} & = & \frac{1}{2} C_a K \equiv \frac{1}{2} C_a \left[ C_A \left( \frac{67}{18} - \frac{\pi^2}{6} \right) - \frac{10}{9} T_R N_f \right], \label{a2number}
\end{eqnarray}
where $C_q = C_F, \, C_g = C_A$, $N_f$ is the number of fermions, and $T_R$ determines the normalization of the generators of fermion representation $R$, $T_F = \frac{1}{2},\, T_A = N_c$, with $N_c$ the number of colors. We will now apply our method to the recalculation of these coefficients.

The only web at order 1 in Feynman gauge is a single gluon exchanged between the two eikonal lines. The color weight is  $\overline{C}^{(1)} = C_a$ by definition (\ref{colwe}), and the web has no internal momenta. Straightforwardly we obtain
\be
2 \int  \frac{d^n k}{(2 \pi)^n} W_{aa}^{(1)}  =  \overline{C}^{(1)}  \left( \frac{\alpha_s(\mu^2)}{\pi} \right) \left( \frac{\mu^2}{m^2} \right)^\varepsilon (4 \pi)^\varepsilon\; \Gamma(\varepsilon)  \int\limits_0^\infty \frac{d k^+}{k^+}.  \label{1loopcalc}
\ee
So at lowest order we get from Eq. (\ref{aendeq})  $A_a^{(1)} = \overline{C}^{(1)} = C_a$, as in (\ref{a1number}).

At order 2 we have the webs shown in Fig. \ref{2loops}, where we rotated the eikonal lines in the figure compared to the diagrams shown in Section \ref{sectexp}, to make the connection to Figs. \ref{partondef} a) and \ref{partonfactend} more evident. The original color factors are (compare to Fig. \ref{order2})
\ba
C(W_b) & = & \left(C_F - \frac{C_A}{2} \right) C_a, \nonumber \\
C(W_c) & = & C(W_d) = - \frac{C_A}{2} C_a \label{2loopcolwe}
\ea
for eikonal lines in the $a$-representation. The respective color weights of the webs c) and d) are the same as the original color factors, since they do not have decompositions (these diagrams are ``maximally nonabelian''). The decomposition of diagram b) was shown as an example in Section \ref{examplesubsect}, which resulted in a color weight
\be
\bar{C}(W_b) =  - \frac{C_A}{2} C_a. \label{colweb}
\ee

The contribution of Fig. \ref{2loops} a) is easily found from Eq. (\ref{1loopcalc}) and the well-known finite terms (see e.\,g. \cite{bookTASI}) after renormalization of the of the gluon-self energy in the $\MS$ scheme, which is an example of what we called a QCD renormalization in the previous subsection:
\be
A_a^{(2),\,a)} = \frac{29}{36} C_A C_a + \frac{1}{18} C_A C_a - \frac{5}{9} T_R N_f C_a = \left(\frac{31}{36} C_A -
\frac{5}{9} T_R N_f \right) C_a. \label{2loopgluon}
\ee
 The first term in the first equality in Eq. (\ref{2loopgluon}) stems from the gluon loop, the second term from the ghost loop. The last term is obviously the fermion loop contribution found from the expression for the graph,
\be
 2 \int  \frac{d^n k}{(2 \pi)^n} W_{aa,\,a)N_f}^{(2)}  = - T_R N_f C_a  \left( \frac{\alpha_s(\mu^2)}{\pi} \right)^2  \left( \frac{\mu^2}{m^2} \right)^{2 \varepsilon} (4 \pi)^{2\varepsilon} \frac{\Gamma(2 \varepsilon)}{\varepsilon} 2 B(2-\varepsilon,2-\varepsilon) \int\limits_0^\infty \frac{d k^+}{k^+}, \label{bubblenf}
\ee
and its counterterm.

\begin{figure}[hbt]
\begin{center}
\epsfig{file=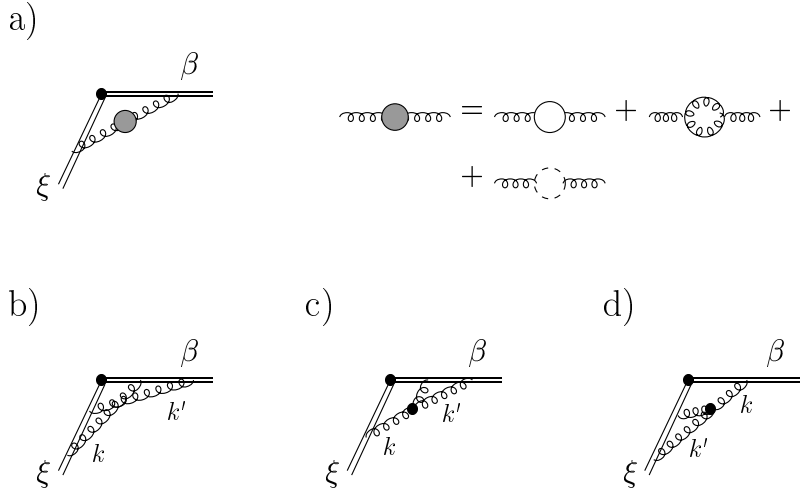,height=6.5cm,clip=0}
\caption{Webs contributing to $A^{(2)}_f$ (compare to Fig. \ref{expgraph}): a) web of order 1 with 1-loop gluon self-energy inserted, b) the ``crossed ladder", c) and d) graphs with a triple gluon vertex.} \label{2loops}
\end{center}
\end{figure}

The LCOPT diagrams obtained from the webs \ref{2loops} b)-d) are shown in Fig. \ref{2loopfig}. We see that due to the numerator $(2 k'^- - k^-)$ in the triple-gluon vertex, web d) contains two orderings on the light-cone; the factors of $2$ and $(-1)$ next to the eikonal vertices in the figure come from this numerator. Furthermore,  for web b) it is important to route the momenta in the crossed ladder independently of each other, such that both of them flow through the vertex, to separate the subdivergence associated with the upper loop ($k'$) from the overall UV divergence.

\begin{figure}
\begin{center}
\epsfig{file=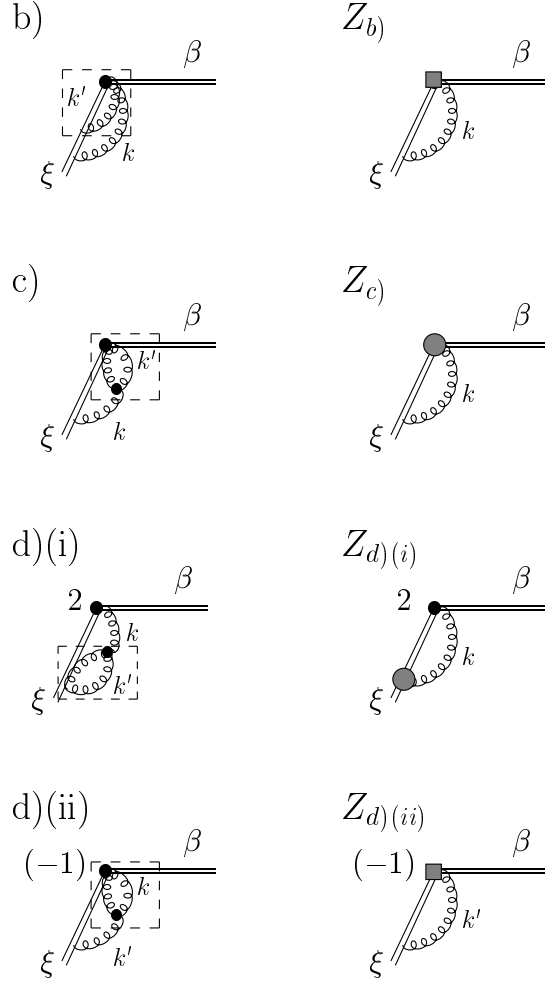,height=13cm,clip=0}
\caption{LCOPT diagrams obtained from Fig. \ref{2loops} b)-d). The subgraphs in the dashed boxes are UV divergent 1-loop subgraphs, whose counterterms are shown in the second column. The grey boxes denote the eikonal vertex counterterms, whereas the grey blobs are the triple-gluon-vertex counterterms, shown in Fig. \ref{3gcounter}. }  \label{2loopfig}
\end{center}
\end{figure}

Now we determine the divergent 1-loop subgraphs for each web by naively counting the powers of transverse momentum components in numerators and denominators. The UV divergent subgraphs are marked with boxes in Fig. \ref{2loopfig}. We see that for web c) and the first term of web d) we need a QCD counterterm for the triple-gluon vertex, whereas for the webs a) and d)(ii) we require vertex counterterms, as shown in the second column of Fig. \ref{2loopfig}. Web d)(ii) is an example for a LCOPT graph with a triple-gluon vertex which does not need QCD renormalization, in contrast to loop-corrections to 3-gluon-vertices in every covariant diagram. Due to the factor of (-1) in web d)(ii) the two vertex counterterms cancel each other, as announced above. The QCD counterterm, as shown in Fig. \ref{3gcounter}, is in the $\MS$ scheme for quark eikonal lines $\beta$ given by
\be
Z^{a\,\mu}_{\mbox{\tiny 3-g},\,ij} = - \frac{C_A}{2} T^a_{ij} \frac{\alpha_s}{\pi} \, g \, \beta^{\mu}  \frac{1}{2} \left( \frac{1}{\varepsilon} - \ln \frac{e^{\gamma_E}}{4 \pi}  \right), \label{3gcount}
\ee
where $g$ is the QCD coupling, $\alpha_s = g^2/(4\pi)$, and $\varepsilon > 0$.

The next step, after adding the appropriate counterterms to the respective graphs, is to perform the plus-momentum integrals. To do so, we dimensionally continue to $\varepsilon < 0$, that is, to $n > 4$ dimensions. The results for the webs b)-d) and the counterterms for UV subdivergences, denoted by $Z$ (omitting the vertex counterterms which cancel each other) are:
\ba
2 \int  \frac{d^n k}{(2 \pi)^n} W_{aa,\, b)}^{(2)} & = & - \overline{C}^{(2)} \left( \frac{\alpha_s(\mu^2)}{\pi} \right)^2 \left(\frac{\mu^2}{m^2}\right)^{2 \varepsilon} (4 \pi)^{2 \varepsilon} \int\limits_0^\infty \frac{d k^+}{k^+} \, \frac{1}{2}\, \frac{\Gamma(2 \varepsilon)}{\varepsilon} B(1+\varepsilon,-\varepsilon),  \label{twoloopa} \\
2 \int  \frac{d^n k}{(2 \pi)^n} W_{aa,\, c)}^{(2)} & = & 2 \int  \frac{d^n k}{(2 \pi)^n} W_{aa,\, d)}^{(2)} = \overline{C}^{(2)} \left( \frac{\alpha_s(\mu^2)}{\pi} \right)^2  \left(\frac{\mu^2}{m^2}\right)^{2 \varepsilon} (4 \pi)^{2 \varepsilon} \int\limits_0^\infty \frac{d k^+}{k^+} \, \frac{1}{4} \frac{\Gamma(2 \varepsilon)}{\varepsilon} \nonumber \\
& & \qquad \qquad \qquad \qquad \qquad \times \, \left\{ B(1-\varepsilon,-\varepsilon) - 2 B(1 - \varepsilon,1-\varepsilon)\right\}, \\
2 \int  \frac{d^n k}{(2 \pi)^n} Z_{c)} & = & 2 \int  \frac{d^n k}{(2 \pi)^n} Z_{d)(i)} = \overline{C}^{(2)} \left( \frac{\alpha_s(\mu^2)}{\pi} \right)^2 \left(\frac{\mu^2}{m^2}\right)^{\varepsilon} (4 \pi)^{\varepsilon} \int\limits_0^\infty \frac{d k^+}{k^+} \,\frac{1}{2}\,  \Gamma( \varepsilon) \left( \frac{1}{\varepsilon} - \ln \frac{e^{\gamma_E}}{4 \pi} \right). \nonumber \\
& & \label{twoloopz}
\ea
 The color weight, as stated in Eqs. (\ref{2loopcolwe}) and (\ref{colweb}), is $\overline{C}^{(2)} = - \frac{C_A}{2} C_a$ for all diagrams. We notice that diagram d) gives the same contribution as its upside-down counterpart c), as expected, but only after adding different types of counterterms.

\begin{figure}[hbt]
\begin{center}
\epsfig{file=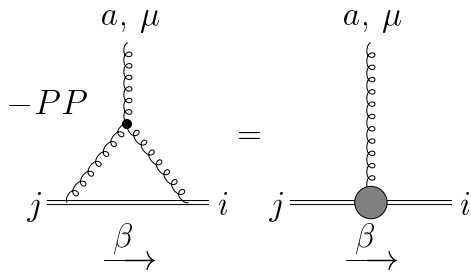,height=3.5cm,clip=0}
\caption{QCD counterterm for the triple-gluon vertex, where $PP$ denotes the pole part (omitting scheme-dependent constants).}  \label{3gcounter}
\end{center}
\end{figure}

After summing over the contribution of the webs b)-d) and the counterterms, we see that the  infrared poles $1/(-\varepsilon)$ in the Beta-functions cancel, as well as the vertex counterterms, leaving us with
\be
A_a^{(2),\,b)-d)} = \frac{C_A}{2} C_a \left( 2 - \frac{\pi^2}{6}\right) \label{remaincont}
\ee
according to Eq. (\ref{aendeq}).
The contributions of all diagrams, (\ref{2loopgluon}) and (\ref{remaincont}), result in the 2-loop coefficient (\ref{a2number}), as announced.

\section{Higher Loops} \label{sect3loop}

\subsection{$N_f^{n-1}$-Terms in $A^{(n)}$}

It is relatively straightforward to obtain a general formula for the $N_f^{n-1}$-contribution to the $n$-loop coefficient $A^{(n)}$, since the only graphs involved are one-loop webs with $n-m-1$ fermion bubbles and $m$ counterterms for the fermion bubbles inserted into the gluon propagator. It is therefore a matter of simple combinatorics to obtain the $\alpha_s^n N_f^{n-1}$ contribution (compare to the one-loop expression Eq. (\ref{bubblenf}) ):
\ba
2 \int \frac{d^{4-2 \varepsilon} k}{(2 \pi)^{4-2 \varepsilon}} W_{aa,\, N_f^{n-1}}^{(n)} & =  & 2\; C_a T_R^{n-1} N_f^{n-1}  \left( \frac{\alpha_s(\mu^2)}{\pi} \right)^n \int\limits_0^\infty \frac{d k^+}{k^+} \nonumber \\
& \times & \sum\limits_{m=0}^{n-1} \left( \begin{array}{c} n-1 \\ m \end{array} \right) (-1)^{n-m-1}\left(\frac{\mu^2}{m^2}\right)^{(n-m)  \varepsilon}
   (4 \pi)^{(n-m) \varepsilon} \; 2^{n-m-2}
\\ & \times & \frac{\Gamma\left( (n-m) \varepsilon \right)}{\Gamma\left( 1+(n-m-1) \varepsilon \right)}
\left[ \Gamma(\varepsilon) B(2-\varepsilon,2-\varepsilon) \right]^{n-m-1} \left[ \frac{1}{3}
\left( \frac{1}{\varepsilon} - \ln \frac{e^{\gamma_E}}{4 \pi} \right) \right]^m . \nonumber
\ea
The $\frac{1}{\varepsilon}$-pole in the expansion of the $\Gamma$- and Beta-functions in the sum is the contribution to the anomalous dimension (cf. Eq. (\ref{aendeq}) ). The contributions up to $\alpha_s^6$ are given in Table \ref{table}. They coincide with the corresponding values (the $\ln N$-terms, or equivalently, the $S_1(N)$-terms) calculated by Gracey in \cite{Gracey}\footnote{Note the different overall normalization of the anomalous dimension there.}.

\begin{table}
\begin{center}
\begin{tabular}{|c|l|}
\hline
$n$ & $C_a (T_R N_f)^{n-1}$-term in $A_a^{(n)}$  \\ \hline
2 &  $-\frac{5}{9}$ \\
3 &  $-\frac{1}{27}$  \\
4 &  $-\frac{1}{81} + \frac{2}{27}  \zeta(3)$ \\
5 &  $-\frac{1}{243} - \frac{10}{243} \zeta(3) + \frac{\pi^4}{2430}$ \\
6 &  $-\frac{1}{729} - \frac{2}{729} \zeta(3) - \frac{\pi^4}{4374} + \frac{2}{81} \zeta(5)$  \\
\hline
\end{tabular}
\end{center}
\caption{$\alpha_s(\mu^2)^n N_f^{n-1} \left[ \frac{1}{1-x} \right]_+$-contributions to the anomalous dimension $P_{ff}$. The expansion of $A_f$ is performed in terms of $\alpha_s/\pi$ (cf. Eq. (\ref{aaa}) ).} \label{table}
\end{table}

\subsection{Towards the Three-Loop Coefficient $A^{(3)}_f$}

Vogt \cite{vogt} obtained a numerical parametrization of the $A^{(3)}$ from the known integer moments of the splitting function\footnote{Note the expansion in $\left(\frac{\alpha_s}{4 \pi}\right)$ there, whereas we expand in terms of $\left(\frac{\alpha_s}{\pi}\right)$ - see Eq. (\ref{aaa}).}:
\be
A^{(3)}_f = \left[(13.81 \pm 0.14) - (4.31 \pm 0.02) T_R N_f - \frac{1}{27}T_R^2 N_f^2\right]\, C_F. \label{vogtresult}
\ee
We obtained the term proportional to $N_f^2$ in the previous subsection, as listed in Table \ref{table}. Now we will go on to compute the term proportional to $N_f$. All intermediate expressions are simple enough to be handled by the general algebraic computer program \textit{Mathematica} \cite{mathematica}. For the calculation of the full $A^{(3)}$ or even higher loops, however, an implementation of the algorithm into a more specialized computer algebra program such as FORM \cite{form} may be desirable.

The diagrams contributing to this term and their QCD counterterms are listed in Table \ref{graphtab}, labelled in analogy to the two-loop case. We only have to compute the contributions from the $g_{\mu \nu}$ part of the dressed gluon propagator, since the longitudinal parts $\sim k_\mu k_\nu$ cancel due to the Ward identity shown in Fig. \ref{subfact}. This cancellation has been verified explicitly.

The contributions
to the set a) are easily computed to be
\be
A_f^{(3),\,a)} = \frac{1}{18} C_A T_R N_f C_F. \label{A3a}
\ee
The contributions to the $N_f$-part of the two-loop gluon self-energy inserted into a one-loop web (set g) ) give:
\be
A_f^{(3),\,g)} = - \left[ C_A \left( \frac{509}{864} + \frac{1}{2} \zeta(3) \right) - C_F \left( - \frac{55}{48}+ \zeta(3) \right)\right] T_R N_f C_F. \label{A3g}
\ee
To compute the two-loop gluon self-energy, the occurring tensor integrals have been reduced to simple scalar one- and two-loop master integrals using the relations derived in \cite{relation}.
We checked our calculations of the set g) against previous computations of the two-loop gluon self-energy in Feynman gauge, see for example \cite{davydbraat}. Note that this contribution has a term $\sim C_F^2$, which is not ``maximally non-abelian''. The results of \cite{davydbraat} include the longitudinal terms of the gluon propagator, which is dressed with a fermion bubble. Since these terms in the two-loop gluon self-energy, as stated above, cancel against the longitudinal parts in the remaining webs, Eq. (\ref{A3g}) does not contain these contributions.

The expressions for the two-loop webs with a one-loop bubble-insertion are found easily from the corresponding two-loop expressions Eqs. (\ref{twoloopa})-(\ref{twoloopz}), taking into account the proper multiples of $\varepsilon$ in the Gamma- and Beta-functions due to the bubbles.
The calculation of the triangles e) and f) is a bit more nontrivial. The resulting contributions can be found in the table. The results for e) and f) have been expanded in terms of $\varepsilon$ and Beta-functions using various identities tabulated in \cite{polylog}.

Since the infrared structure of the graphs is modified by the bubbles, which effectively raise the powers of the corresponding gluon propagators by $\varepsilon$ to a non-integer value, the upside-down counterparts do not give the same contributions. This asymmetry is not surprising, since we compute the coefficients collinear to the plus eikonal, thus introducing an asymmetry in how we treat the eikonal lines and the gluons attaching to them. However, we find that the sum of graphs in set d) gives the same contribution as the sum of graphs in set c), as can be seen from the tabulated expressions.

The individual diagrams b)-f) have at most three UV (QCD) divergences and one IR/collinear divergence, in addition to the overall scaleless $k^+$-integral.
We observe that the diagrams
with a one-loop counterterm for the fermion bubble and the one-loop counterterms for the triangle graphs have the same IR structure as the two-loop webs. Thus their IR divergences cancel separately from the rest of the diagrams. This implies that the collinear divergences have to cancel within the set of remaining diagrams, that is, within the set of webs with bubbles and the triangles. Moreover, we observe that the infrared divergences cancel within certain subsets of these graphs. Namely, they cancel separately between graphs b)(1), c)(1), and d)(1), between graphs b)(2), c)(2), and d)(2), as well as between c)(3), d)(3), e) and f).

Summing over all contributions from graphs b)-f) we arrive at
\be
A_f^{(3),\,b)-f)} = - \left(\frac{125}{288} - \frac{5 \pi^2}{54} + \frac{2 \zeta(3)}{3} \right) T_R N_f C_A C_F. \label{A3b}
\ee

We performed several checks of our computations. The infrared structure described above is one check of the results listed in Table \ref{graphtab}. Another check is the cancellation of non-local logarithms $\sim \log M$. Furthermore, the values of the $1/\varepsilon^3$- and $1/\varepsilon^2$-poles can be predicted from the one- and two-loop calculations performed in Section \ref{sectexample} \cite{tHooft}. The sum of all diagrams contributing at $\alpha_s^3 N_f$ has the following structure:
\ba
2 \int \frac{d^n k}{(2\pi)^n} W^{(3)}_{aa,\,N_f} & = & \left\{ - \frac{11}{54} C_A C_F T_R   \frac{1}{\varepsilon^3} + \right. \label{poles} \\
& & +\, \left.
 \left[ \left( \frac{167}{324} -\frac{\pi^2}{108} \right) C_A + \frac{1}{12} C_F
  \right] C_F T_R  \frac{1}{\varepsilon^2}
 +  \frac{1}{3}  A_{N_f}^{(3)} \frac{1}{\varepsilon}   \right\}  N_f \left( \frac{\alpha_s}{\pi} \right)^3 \int\limits_0^\infty \frac{d k^+}{k^+}. \nonumber
\ea
 The predictions of the higher poles in Eq. (\ref{poles})  coincide with the poles obtained from the expansion of the calculated expressions listed in the table.

Adding (\ref{A3a}), (\ref{A3g}), and (\ref{A3b}) we obtain the term proportional to $N_f$ contributing to the three-loop coefficient $A^{(3)}$:
\be
A^{(3)}_{N_f} =  - \left[ C_A \left( \frac{209}{216} - \frac{5 \pi^2}{54} + \frac{7 \zeta(3)}{6} \right)  -  C_F \left(- \frac{55}{48}+ \zeta(3) \right)\right] T_R N_f C_F  = -4.293\, T_R N_f C_F,
\ee
which agrees with the numerical prediction in Eq. (\ref{vogtresult}).

\begin{table}[hbt]
 Table \ref{graphtab}:
\begin{center}
\begin{tabular}{|l|c|l|}
\hline
\hspace*{1cm} Web & Factor  & \hspace*{1cm} Contribution \\ \hline
 & & \\
\raisebox{3mm}{a)} \raisebox{-3mm}{ \epsfig{file=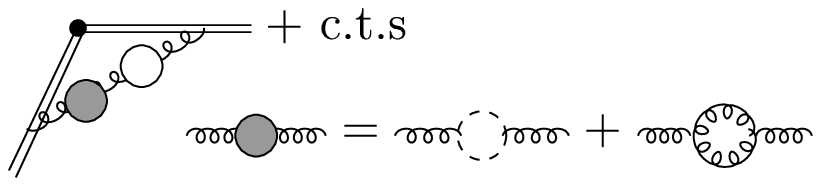,height=1cm,clip=0}}   & 2  & see Eq. (\ref{A3a}) \\
 & & \\ \hline\hline
 & & \\
 \raisebox{3mm}{b)(1)}\hspace*{4mm} \raisebox{-3mm}{ \epsfig{file=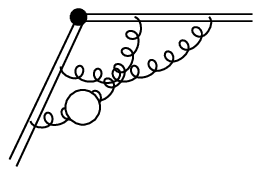,height=1cm,clip=0}} & 2 & $- K M^{3 \varepsilon} \frac{\Gamma(3 \varepsilon)}{2 \varepsilon^2} B(2-\varepsilon,2-\varepsilon) B(1+\varepsilon,-\varepsilon)$ \\
 & & \\
 \raisebox{3mm}{b)(2)}\hspace*{4mm} \raisebox{-3mm}{ \epsfig{file=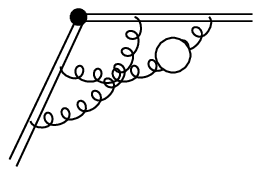,height=1cm,clip=0}} & 2 & $- K M^{3 \varepsilon} \frac{\Gamma(3 \varepsilon)}{4 \varepsilon^2} B(2-\varepsilon,2-\varepsilon)  B(1+2\varepsilon,-2\varepsilon)$ \\
 & & \\ \hline
 & & \\
 \raisebox{3mm}{b)(C1)}\hspace*{1.8mm} \raisebox{-3mm}{ \epsfig{file=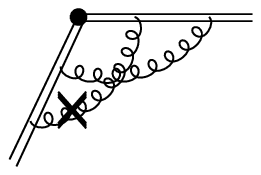,height=1cm,clip=0}} & 4 & $+ K M^{2 \varepsilon} \frac{\Gamma(2 \varepsilon)}{12 \varepsilon} N_{\varepsilon} B(1+\varepsilon,-\varepsilon)$ \\
 & & \\
\hline\hline
 & &\\
 \raisebox{3mm}{c)(1)}\hspace*{4mm} \raisebox{-3mm}{ \epsfig{file=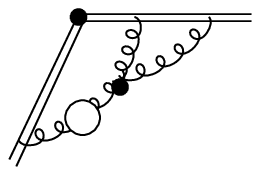,height=1cm,clip=0}} & 2 & $+ K M^{3 \varepsilon} \frac{1}{4} \frac{\Gamma(3 \varepsilon)}{e^2} \frac{\left(\Gamma(1+\varepsilon)\right)^2}{\Gamma(1+2 \varepsilon)} B(2-\varepsilon,2-\varepsilon)$ \\
 & & $\qquad \times \left\{ B(1-\varepsilon,-\varepsilon) - 2 B(1 - \varepsilon,1-\varepsilon)\right\}$ \\
 & & \\
  \raisebox{3mm}{c)(2)}\hspace*{4mm} \raisebox{-3mm}{ \epsfig{file=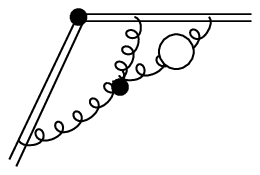,height=1cm,clip=0}} & 2 & $+ K M^{3 \varepsilon} \frac{1}{4} \frac{\Gamma(3 \varepsilon)}{2 \varepsilon^2} B(2-\varepsilon,2-\varepsilon)$ \\
 & & $\qquad \times \left\{ B(1-\varepsilon,-2 \varepsilon) - 2 B(1 - \varepsilon,1-2\varepsilon)\right\}$ \\
 & & \\
 \raisebox{3mm}{c)(3)}\hspace*{1.8mm} \raisebox{-3mm}{ \epsfig{file=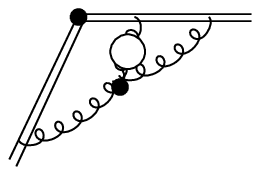,height=1cm,clip=0}} & 2 & $+ K M^{3 \varepsilon} \frac{1}{4} \frac{\Gamma(3 \varepsilon)}{2 \varepsilon^2} B(2-\varepsilon,2-\varepsilon)$ \\
 & & $\qquad \times \left\{ B(1-2\varepsilon,- \varepsilon) - 2 B(1 - \varepsilon,1-2\varepsilon)\right\}$ \\
\hline
 & &\\
 \raisebox{3mm}{d)(1)}\hspace*{4mm} \raisebox{-3mm}{ \epsfig{file=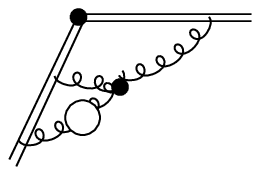,height=1cm,clip=0}} & 2 & $+ K M^{3 \varepsilon} \frac{1}{4} \frac{\Gamma(3 \varepsilon)}{\varepsilon^2}  B(2-\varepsilon,2-\varepsilon)$ \\
 & & $\qquad \times \left\{ \frac{\left(\Gamma(1+\varepsilon)\right)^2}{\Gamma(1+2 \varepsilon)}  B(1-\varepsilon,-\varepsilon) - B(1 - 2 \varepsilon,1-\varepsilon)\right\}$ \\
 & & \\
  \raisebox{3mm}{d)(2)}\hspace*{4mm} \raisebox{-3mm}{ \epsfig{file=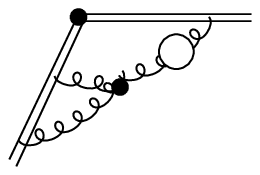,height=1cm,clip=0}} & 2 & $+ K M^{3 \varepsilon} \frac{1}{4} \frac{\Gamma(3 \varepsilon)}{2 \varepsilon^2} B(2-\varepsilon,2-\varepsilon)$ \\
 & & $\qquad \times \left\{ B(1-\varepsilon,-2 \varepsilon) - 4  \frac{\left(\Gamma(1+\varepsilon)\right)^2}{\Gamma(1+2 \varepsilon)} B(1 - \varepsilon,1-2\varepsilon)\right\}$ \\
 & & \\
 \raisebox{3mm}{d)(3)}\hspace*{1.8mm} \raisebox{-3mm}{ \epsfig{file=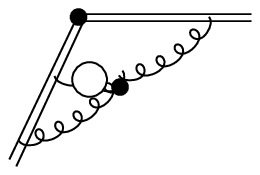,height=1cm,clip=0}} & 2 & $+ K M^{3 \varepsilon} \frac{1}{4} \frac{\Gamma(3 \varepsilon)}{2 \varepsilon^2} B(2-\varepsilon,2-\varepsilon)$ \\
 & & $\qquad \times \left\{ B(1-2\varepsilon,- \varepsilon) - 2 B(1 - \varepsilon,1-2\varepsilon)\right\}$ \\
\hline
\end{tabular}
\end{center}
\end{table}

\begin{table}
Continuation of Table \ref{graphtab}:
\begin{center}
\begin{tabular}{|l|c|l|} \hline
\hspace*{1cm} Web & Factor  & Contribution  \\ \hline
 &  &\\
\raisebox{3mm}{c)d)(C1)}\raisebox{-3mm}{ \epsfig{file=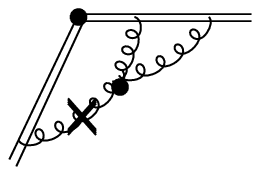,height=1cm,clip=0}} & 12 & $- K M^{2 \varepsilon} \frac{1}{24} \frac{\Gamma(2 \varepsilon)}{ \varepsilon}  N_{\varepsilon} \left\{ B(1-\varepsilon,-\varepsilon) - 2 B(1 - \varepsilon,1-\varepsilon)\right\}$ \\
 & & \\
\raisebox{3mm}{c)d)(C2)}\raisebox{-3mm}{ \epsfig{file=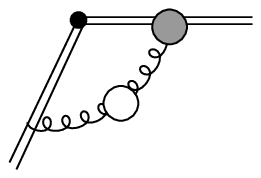,height=1cm,clip=0}} & 4 & $+K M^{2 \varepsilon} \frac{\Gamma(2 \varepsilon)}{2 \varepsilon} N_{\varepsilon} B(2-\varepsilon,2-\varepsilon) $ \\
 & & \\
\hline
 & & \\
\raisebox{3mm}{c)d)(C3)}\raisebox{-3mm}{ \epsfig{file=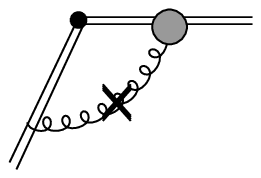,height=1cm,clip=0}} & 4 & $-K M^{ \varepsilon} \frac{\Gamma(\varepsilon)}{12} N_{\varepsilon}^2$ \\
 & & \\
\raisebox{3mm}{c)d)(C4)}\raisebox{-3mm}{ \epsfig{file=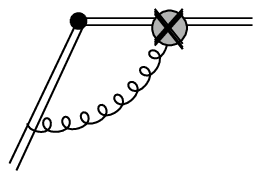,height=1cm,clip=0}} & 8 & $-K M^\varepsilon \frac{\Gamma(\varepsilon)}{2} \left[ \frac{1}{12}  N_{\varepsilon}^2 - \frac{1}{18} N_{\varepsilon} \right] $ \\
 & & \\
\hline\hline
 & & \\
\raisebox{3mm}{e)} \hspace*{4mm} \raisebox{-3mm}{ \epsfig{file=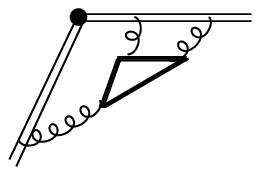,height=1cm,clip=0}} & 2 & $- K M^{3 \varepsilon} \frac{\Gamma(3 \varepsilon)}{8 \varepsilon^2 } B(2-\varepsilon,2-\varepsilon) \left\{ B(3-2\varepsilon,-\varepsilon) \right. $ \\
 & & $\quad \left. - B(1-\varepsilon,2-2\varepsilon) + \frac{2 \pi^2}{3} \varepsilon + \left(4 \zeta(3)-2\right) \varepsilon^2  \right\} $ \\
\hline
 & & \\
\raisebox{3mm}{f)} \hspace*{4mm} \raisebox{-3mm}{ \epsfig{file=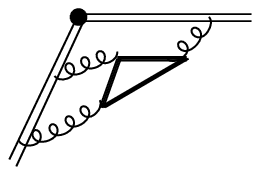,height=1cm,clip=0}} & 2 & $- K M^{3 \varepsilon} \frac{\Gamma(3 \varepsilon)}{8 \varepsilon^2} B(2-\varepsilon,2-\varepsilon)\left\{  B(3-2\varepsilon,-\varepsilon)\right. $ \\
 & & $\quad \left. - B(1-\varepsilon,2-2\varepsilon) - \frac{\pi^2}{3} \varepsilon + \left(10 \zeta(3) -2 \right)  \varepsilon^2 \right\}$ \\
\hline
 & & \\
\raisebox{3mm}{e)f)(C1)}\raisebox{-3mm}{ \epsfig{file=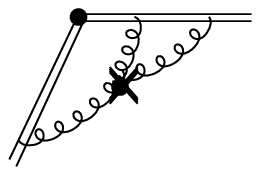,height=1cm,clip=0}} & 4 & $+ K M^{2 \varepsilon} \frac{1}{24} \frac{\Gamma(2 \varepsilon)}{ \varepsilon} N_{\varepsilon} \left\{ B(1-\varepsilon,-\varepsilon) - 2 B(1 - \varepsilon,1-\varepsilon)\right\} $ \\
 & & \\ \hline
 & & \\
\raisebox{3mm}{e)f)(C2)}\raisebox{-3mm}{ \epsfig{file=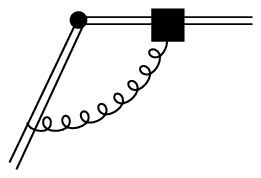,height=1cm,clip=0}} & 4 & $+ K M^\varepsilon \frac{\Gamma(\varepsilon)}{2} \frac{1}{12} \left( N_{\varepsilon}^2 - \frac{5}{12}  N_{\varepsilon}  \right)$  \\
 & & \\ \hline \hline
\end{tabular}
\end{center}
\end{table}

\begin{table}
Continuation of Table \ref{graphtab}:
\begin{center}
\begin{tabular}{|l|c|l|} \hline
\hspace*{1cm} Web & Factor  & Contribution  \\ \hline
& & \\
\raisebox{3mm}{g) }\raisebox{-3mm}{ \epsfig{file=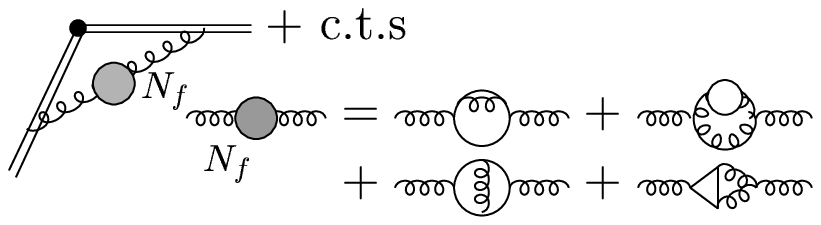,height=1.3cm,clip=0}} & 2 & see Eq. (\ref{A3g}) \hspace*{2cm} \\
 & & \\
\hline\hline
\end{tabular}
\end{center}
We introduced the following abbreviations:
\ba
K & \equiv & \frac{C_A}{2} T_R N_f C_F \left( \frac{\alpha_s}{\pi}\right)^3 \int\limits_0^\infty \frac{dk^+}{k^+}, \\
M & \equiv & \left(\frac{\mu^2}{m^2} \right) \left(4 \pi \right), \\
N_{\varepsilon} & \equiv & \frac{1}{\varepsilon} - \ln \frac{e^{\gamma_E}}{4 \pi} .
\ea
\caption{Webs contributing to the $N_f$-term of the three-loop coefficient $A^{(3)}$ and their counterterms (c.t.s), labelled (C). The cross denotes the counterterm for the fermion bubble. Similarly, the cross in the triple-gluon vertex denotes the counterterm for the fermion triangle. The grey blob represents the counterterm Fig. \ref{3gcounter}, the grey blob with a cross is the 2-loop counterterm for the triple-gluon vertex with a fermion bubble inside. And finally, the black box denotes the fermion part of the 2-loop counterterm for the triple-gluon vertex.
 We omitted vertex counterterms which cancel. We refrain from drawing all counterterms which give the same contribution. Instead, we indicate multiple contributions in the column ``factor''. A factor of 2 is due to the two complex conjugate contributions, and has already been taken into account in Eqs. (\ref{A3a}) and (\ref{A3g}). } \label{graphtab}
\end{table}

\clearpage

\section{Conclusions}

We have developed and proved a method for the calculation of the coefficients proportional to $\left[\frac{1}{1-x}\right]_+$ of the non-singlet parton splitting functions, whose knowledge, for example, is important for NNLL resummations. The method is based on the factorization properties of the splitting functions, and on the exponentiation of eikonal cross sections.

We illustrated the method with the rederivation of the 1- and 2-loop coefficients $A^{(1)}$ and $A^{(2)}$, as well as the $N_f^{n-1}$ terms at order $n$. We presented the result for the term proportional to $N_f$ at three loops, which coincides with the approximate result obtained by Vogt \cite{vogt}. The full splitting functions at three loops are currently being computed by Moch, Vermaseren, and Vogt \cite{MoVe} with the help of the operator product expansion.  Their results for the $N_f$-term \cite{MVV} provide a further check of our calculations.

Although a calculation via the OPE provides the complete $N$, or equivalently, $x$-dependence of the splitting functions, it involves a large number of diagrams and complex expressions at higher orders. A computation at three loops is a formidable task, and it seems unlikely that higher loop calculations will be completed in the near future. However, for certain observables, large logarithms due to soft and/or collinear radiation become numerically important, and need to be resummed to as high a level in logarithms as possible. Our method, although limited to only the computation of $A$, has the advantage that higher-order computations are much less complex than within conventional methods, because the number of graphs is greatly reduced, and the expressions involved are relatively simple in LCOPT. Moreover, a fully computerized implementation of the algorithm should be straightforward. Therefore, the computation of the coefficients $A$ at four or even higher loops may be within reach.

\begin{appendix}

\section{Rules for LCOPT} \label{Lrules}

As already stated above, these rules can be obtained by  performing first all minus integrals
\cite{changma, kosop}.
\begin{itemize}
\item We start with forming all possible light-cone time orderings of a given covariant diagram.
\item Only those configurations are kept which describe possible physical processes once the energy flow is specified.
\item For every loop we have a factor
\be
\frac{d l^+\,d^{2-2\varepsilon} l_{\perp}}{(2 \pi)^{3 -2 \varepsilon}}
\ee
 if we work in $n = 4 - 2 \varepsilon$ dimensions.
\item For every internal line we have a factor
\be
\frac{\theta(l_i^+)}{2 l_i^+},
\ee
 corresponding to the flow of plus momentum through the graph.
\item Every intermediate virtual state contributes a factor
\be
\frac{i}{q^- - \sum_j \frac{l^2_{j,\,\perp}}{2 l_j^+} + i \epsilon},  \label{denom}
\ee
where we sum over all momenta comprising that virtual state, and where $q^-$ is the external minus-momentum of the incoming state(s).
\item Every intermediate real state gives a momentum conserving delta-function:
\be
2 \pi \delta \left(q^- - \sum_j \frac{l^2_{j,\,\perp}}{2 l_j^+} \right),
\ee
where the sum is over all momenta in that real state.
\item Since all lines are on-shell, we replace for every Fermion numerator its minus component by its on-shell value:
\be
l^- = \frac{l_\perp^2}{2 l^+}.
\ee
\end{itemize}

\end{appendix}

\subsection*{Acknowledgements}
I am deeply indebted to George Sterman for his advice and his support in disentangling the web of complications encountered during the production of this paper. I am very grateful to Maria Elena Tejeda-Yeomans for her advice and for sharing her expertise in calculational techniques. I thank Jos Vermaseren and Andreas Vogt for extremely helpful communication. Also, I appreciate valuable conversations with Lilia Anguelova, John A. Gracey, Tibor K\'ucs, Jack Smith, and Willy van Neerven. This work was supported in part by the National Science Foundation, grant  PHY-0098527.


\begin{thebibliography}{99}


\bibitem{George} G. Sterman, in \textit{AIP Conference Proceedings Tallahassee, Perturbative Quantum Chromodynamics}, eds. D. W. Duke, J. F. Owens, New York, 1981.


\bibitem{gath} J. G. M. Gatheral, Phys. Lett. \textbf{133 B}, 90 (1983).

\bibitem{freta} J. Frenkel, J. C. Taylor, Nucl. Phys. \textbf{B 246}, 231 (1984).


\bibitem{vogt}
 A. Vogt, Phys. Lett. \textbf{B 497}, 228 (2001) [hep-ph/0010146].\\
 W. L. van Neerven, A. Vogt, Nucl. Phys. \textbf{B 603}, 42 (2001) [hep-ph/0103123].  \\
 W. L. van Neerven, A. Vogt, J. Phys. \textbf{G 28}, 727 (2002) [hep-ph/0107194].


\bibitem{DGLAP} V. N. Gribov, L. N. Lipatov, Sov. J. Nucl. Phys. \textbf{15}, 438 (1972). \\
L. N. Lipatov, Sov. J. Nucl. Phys. \textbf{20}, 95 (1975). \\
 G. Altarelli, G. Parisi, Nucl. Phys. \textbf{B126}, 298 (1977). \\
 Y. L. Dokshitzer, Sov. Phys. JETP \textbf{46}, 641 (1977).

\bibitem{pQCD} J. C. Collins, D. E. Soper, G. Sterman, in \textit{Perturbative Quantum Chromodynamics}, ed. A. H. Mueller, World Scientific, Singapore, 1989, p. 1.


\bibitem{korch} G. P. Korchemsky, Mod. Phys. Lett. \textbf{A 4}, 1257 (1989).

\bibitem{alball} S. Albino, R. D. Ball, Phys. Lett. \textbf{B 513}, 93 (2001) [hep-ph/0011133].


\bibitem{threshold} G. Sterman, Nucl. Phys. \textbf{B281}, 310 (1987). \\
S. Catani, L. Trentadue, Nucl. Phys. \textbf{B 327}, 323 (1989). \\
S. Catani, L. Trentadue, Nucl. Phys. \textbf{B 353}, 183 (1991). \\
N. Kidonakis, Int. J. Mod. Phys. \textbf{A15}, 1245 (2000) [hep-ph/9902484].

\bibitem{vogt2} A. Vogt, in Ref. \cite{vogt}.

\bibitem{2loopknown} D. J. Gross, F. Wilczek, Phys. Rev. D \textbf{8}, 3633 (1973); \textit{ibid.}, D \textbf{9}, 980 (1974).\\
 E. G. Floratos, D. A. Ross, C. T. Sachrajda, Nucl. Phys. \textbf{B 129}, 66 (1977); \textit{ibid.}, \textbf{B 139}, 545 (1978) (erratum). \\
A. Gonzales-Arroyo, C. Lopez, F. J. Yndurain, Nucl. Phys. \textbf{B 153}, 161 (1979). \\
G. Curci, W. Furmanski, R. Petronzio, Nucl. Phys. \textbf{B 175}, 27 (1980). \\
 E. G. Floratos, C. Kounnas, R. Lacaze, Nucl. Phys. \textbf{B 192}, 417 (1981). \\
 J. Kodaira, L. Trentadue, Phys. Lett. \textbf{B 112}, 66 (1982).

\bibitem{Gracey} J. A. Gracey, Phys. Lett. \textbf{B 322}, 141 (1994) [hep-ph/9401214].

\bibitem{3loop} S. Moch, J. A. M. Vermaseren, Nucl. Phys. \textbf{B 573}, 853 (2000) [hep-ph/9912355]. \\
 S. Moch, J. A. M. Vermaseren, Nucl. Phys. \textbf{B 89} (Proc. Suppl.), 137 (2000) [hep-ph/0006053]. \\
   S. Moch, J. A. M. Vermaseren, M. Zhou, hep-ph/0108033.


\bibitem{moments} S. A. Larin, T. van Ritbergen, J. A. M. Vermaseren, Nucl. Phys. \textbf{B427}, 41 (1994). \\
 J. Bl\"umlein, A. Vogt, Phys. Lett. \textbf{B 370}, 149 (1996) [hep-ph/9510410]. \\
 S. A. Larin, P. Nogueira, T. van Ritbergen, J. A. M. Vermaseren, Nucl. Phys. \textbf{B 492}, 338 (1997) [hep-ph/9605317]. \\
 A. R\'etey, J. A. M. Vermaseren, Nucl. Phys. \textbf{B 604}, 281 (2001) [hep-ph/0007294].


\bibitem{MoVe} S. Moch, J. A. M. Vermaseren, hep-ph/0208050. \\
 J. A. M. Vermaseren, private communication.

\bibitem{MVV} S. Moch, J. A. M. Vermaseren, A. Vogt, hep-ph/0209100.


\bibitem{pdfs} J. C. Collins, D. E. Soper, Nucl. Phys. \textbf{B 194}, 445 (1982).


\bibitem{korchmarch} G. P. Korchemsky, G. Marchesini, Nucl. Phys. \textbf{B406}, 225 (1993) [hep-ph/9210281].

\bibitem{cusp} A. M. Polyakov, Nucl. Phys. \textbf{B164}, 171 (1979). \\
 I. Ya. Aref'eva, Phys. Lett. \textbf{B 93}, 347 (1980). \\
 J. Gervais, A. Neveu,  Nucl. Phys. \textbf{B 163}, 189 (1980). \\
  V. S. Dotsenko, S. N. Vergeles, Nucl. Phys. \textbf{B 169}, 527 (1980). \\
 R. A. Brandt, F. Neri, M.-A. Sato, Phys. Rev. \textbf{D 24}, 879 (1981). \\
 G. P. Korchemsky, A. V. Radyushkin, Nucl. Phys. \textbf{B 283}, 342 (1987). \\
 I. A. Korchemskaya, G. P. Korchemsky, Phys. Lett. \textbf{B 287}, 169 (1992).

\bibitem{bookTASI} G. Sterman, \textit{An Introduction to Quantum Field Theory}, Cambridge University Press, Cambridge, 1993. \\
 G. Sterman, in \textit{QCD and Beyond, Proceedings of the Theoretical Advanced Study Institute in Elementary Particle Physics (TASI 95)}, ed. D. E. Soper, World Scientific, Singapore, 1996 [hep-ph/9606312].

\bibitem{count} G. Sterman, Phys. Rev. \textbf{D 17}, 2773 (1978).

\bibitem{landau} L. D. Landau, Nucl. Phys. \textbf{13}, 181 (1959).

\bibitem{glauber} G. T. Bodwin, S. J. Brodsky, G. P. Lepage, Phys. Rev. Lett. \textbf{47}, 1799 (1981).

\bibitem{coste} J. C. Collins, G. Sterman, Nucl. Phys. \textbf{B 185}, 172 (1981).

\bibitem{CSS} J. C. Collins, D. E. Soper, G. Sterman, Nucl. Phys. \textbf{B 261}, 104 (1985). \\
J. C. Collins, D. E. Soper, G. Sterman, Nucl. Phys. \textbf{B 308}, 833 (1988).

\bibitem{back} J. C. Collins, D. E. Soper, Nucl. Phys. \textbf{B 193}, 381 (1981).

\bibitem{GraYen}
G. Grammer, Jr., D. R. Yennie, Phys. Rev. \textbf{D 8}, 4332 (1973). \\
  S. B. Libby, G. Sterman, Phys. Rev. \textbf{D 18}, 3252 (1978).


\bibitem{EGW} E. Laenen, G. Sterman, W. Vogelsang, Phys. Rev. D \textbf{63}, 114018 (2001) [hep-ph/0010080].

\bibitem{levsuch} M. Levy, J. Sucher, Phys. Rev. \textbf{186}, 1656 (1963).


\bibitem{gafreta} J. Frenkel, J. G. M. Gatheral, J. C. Taylor, Nucl. Phys. \textbf{B 233}, 307 (1984).


\bibitem{changma} S.-J. Chang, S.-K. Ma, Phys. Rev. \textbf{180}, 1506 (1969).

\bibitem{kosop} J. B. Kogut, D. E. Soper, Phys. Rev. D \textbf{1}, 2901 (1970).


\bibitem{mathematica} S. Wolfram, \textit{The Mathematica Book}, 4th Edition, Cambridge University Press,  Cambridge, 1999.

\bibitem{form} J. A. M. Vermaseren, \textit{New Features of FORM}, mat-ph/0010025.

\bibitem{relation} G. 't Hooft, M. Veltman, Nucl. Phys. \textbf{B 44}, 189 (1972). \\
F. V. Tkachov, Phys. Lett. \textbf{B 100}, 65 (1981). \\
K. G. Chetyrkin, F. V. Tkachov, Nucl. Phys. \textbf{B 192}, 159 (1981). \\
O. V. Tarasov, Phys. Rev. \textbf{D 54}, 6479 (1996) [hep-ph/9606018]. \\
O. V. Tarasov, Nucl. Phys. \textbf{B 502}, 455 (1997) [hep-ph/9703319].

\bibitem{davydbraat} E. Braaten, J. P. Leveille, Phys. Rev. D \textbf{24}, 1369 (1981). \\
 A. I. Davydychev, P. Osland, O. V. Tarasov, Phys. Rev. D \textbf{58}, 036007 (1998) [hep-ph/9801380].

\bibitem{polylog}
L. Lewin, \textit{Polylogarithms and Associated Functions}, North Holland, Amsterdam, 1981. \\
A. Devoto, D. W. Duke, Riv. Nuovo Cim. \textbf{7N6}, 1 (1984). \\
I. S. Gradshteyn, I. M. Ryzhik, \textit{Table of Integrals, Series, and Products}, 5th Edition, ed. A. Jeffrey,  Academic Press, London, 1994.

\bibitem{tHooft} G. 't Hooft, Nucl. Phys. \textbf{B 61}, 455 (1973).

\end{thebibliography}
\end{document}